\newcommand\Msun{\hbox{M$_\odot$}}
\newcommand\Zsun{\hbox{$Z_\odot$}}
\newcommand\kms{\hbox{km$\,$s$^{-1}$}}

\newcommand\one{\,{\sc i}}
\newcommand\two{\,{\sc ii}}
\newcommand\three{\,{\sc iii}}

\newcommand\tmult{\multicolumn{2}{c}}
\newcommand\hst{HST}

\newcommand\ie{\textit{i.\,e.}}
\newcommand\eg{e.\,g.}
\newcommand\cf{\textit{cf.}}
\newcommand\etal{et~al.}

\newcommand\bvi{\textit{BVI}}

\newcommand\ha{H$\alpha$}
\newcommand\hb{H$\beta$}

\newcommand\n{NGC~}

\newcommand{\farcs}{\mbox{\ensuremath{.\!\!^{\prime\prime}}}}

\newcommand{\isk}{Konstantopoulos}

\documentclass[preprint2]{aastex}

\usepackage{graphicx,epstopdf}
\usepackage{color,soul}
 
\shorttitle{Shocks and Star Formation in Stephan's Quintet I.}
\shortauthors{Konstantopoulos et al.}

\begin{document}

\title{Shocks and Star Formation in Stephan's Quintet.\\I. Gemini Spectroscopy of \ha-bright knots.}

\author{
I.~S.~Konstantopoulos\altaffilmark{1},
P.~N.~Appleton\altaffilmark{2}, 
P.~Guillard\altaffilmark{3}, 
G.~Trancho\altaffilmark{4},
M.~E.~Cluver\altaffilmark{1}, 
N.~Bastian\altaffilmark{5},
J.~C.~Charlton\altaffilmark{6},
K.~Fedotov\altaffilmark{7,8},
S.~C.~Gallagher\altaffilmark{7},
L.~J.~Smith\altaffilmark{9}, 
C.~J.~Struck\altaffilmark{10}
}

\altaffiltext{1}{Australian Astronomical Observatory, PO Box 915, North Ryde NSW 1670, Australia; iraklis@aao.gov.au. ISK is the recipient of a John Stocker Postdoctoral Fellowship from the Science Industry Endowment Fund.}
\altaffiltext{2}{NASA Herschel Science Center (NHSC), California Institute of Technology, Pasadena, CA 91125, USA. }
\altaffiltext{3}{Institut d'Astrophysique Spatiale, Universit\'e Paris-Sud XI, 91405 Orsay Cedex.}
\altaffiltext{4}{Giant Magellan Telescope Organisation, Pasadena, CA 91101, USA.}
\altaffiltext{5}{Astrophysics Research Institute, Liverpool John Moores University, Liverpool L3 5RF, UK.}
\altaffiltext{6}{Department of Astronomy \& Astrophysics, The Pennsylvania State University, University Park, PA 16802, USA.}
\altaffiltext{7}{Department of Physics \& Astronomy, The University of Western Ontario, London, ON, N6A 3K7, Canada.}
\altaffiltext{8}{Herzberg Institute of Astrophysics, Victoria, BC, V9E 2E7, Canada.}
\altaffiltext{9}{Space Telescope Science Institute and European Space Agency, Baltimore, MD 21218, USA.}
\altaffiltext{10}{Department of Physics \& Astronomy, Iowa State University, Ames, IA 50011, USA.}

\begin{abstract}
We present a Gemini-GMOS spectroscopic study of HST-selected \ha-emitting regions in Stephan's Quintet (HCG~92), a nearby compact galaxy group, with the aim of disentangling the processes of shock-induced heating and star formation in its intra-group medium. The $\approx$40 sources are distributed across the system, but most densely concentrated in the $\sim$kpc-long shock region. Their spectra neatly divide them into narrow- and and broad-line emitters, and we decompose the latter into three or more emission peaks corresponding to spatial elements discernible in HST imaging. The emission line ratios of the two populations of \ha-emitters confirm their nature as H\two\ regions (90\% of the sample) or molecular gas heated by a shock-front propagating at $\lesssim$300~\kms. Their redshift distribution reveals interesting three-dimensional structure with respect to gas-phase baryons, with no H\two\ regions associated with shocked gas, no shocked regions in the intruder galaxy \n7318B, and a sharp boundary between shocks and star formation. We conclude that star formation is inhibited substantially, if not entirely, in the shock region. Attributing those H\two\ regions projected against the shock to the intruder, we find a lopsided distribution of star formation in this galaxy, reminiscent of pile-up regions in models of interacting galaxies. The \ha\ luminosities imply mass outputs, star formation rates, and efficiencies similar to nearby star-forming regions. Two large knots are an exception to this, being comparable in stellar output to the prolific 30~Doradus region. We also examine Stephan's Quintet in the context of compact galaxy group evolution, as a paradigm for intermittent star formation histories in the presence of a rich, X-ray emitting intra-group medium.\vspace{10pt}
\end{abstract}

%\keywords{}

%======= SECTION =======
\section{Introduction}\label{sec:intro}
%--------------------------------------
The interplay between shocks and the collapse of gaseous material into stars is as yet not fully understood. While shock fronts are conspicuously observed to propagate through star-forming regions after the first generation of supernova explosions \citep[following the `triggering' paradigm set by][]{elmegreen_lada_77}, one might expect the net effect to be the suppression star formation, as the shock comes with a deposition of energy that should heat up the interstellar medium \citep[such as the strong shocks studied by][]{reynaud98}. An extra complication is the contemporaneous presence of turbulent flows in star-forming regions. At the molecular core level, several models have long advocated turbulence as the dominant collapse mechanism in the interstellar medium \citep[\eg][and many others]{elmegreen02,klessen10,nakamura11}, and have gained support from the recent observations of a correlation between H\one\ line broadening above the thermal value, attributed to turbulence, and the local star formation rate \citep[][based on a sample of $\approx30$ nearby galaxies]{tamburro09}. However, the opposing side argues that turbulence tends to inhibit the process in various ways, \eg, by inducing runaway collapse when acting as the only gravitational agent \citep{hopkins11}, or that such ``local viscosity'' is dwarfed by the effect of global torques \citep{thompson05}. Other models have demonstrated an indifference on the part of star formation to the effects of triggering by turbulence \citep[most notably][]{dale07,getman12}. Finally, while shocks and turbulence are spatially correlated with star formation, an argument was recently made by \citet{mac-low13} against any form of causality: newborn stars may simply be riding in the same carriage as turbulence and shocks. In this scenario, gravitational instabilities are enough to drive star formation, with no need for other input physics. 

\begin{figure*}[phtb]
	\begin{center}
		\includegraphics[width=0.75\linewidth,angle=0]{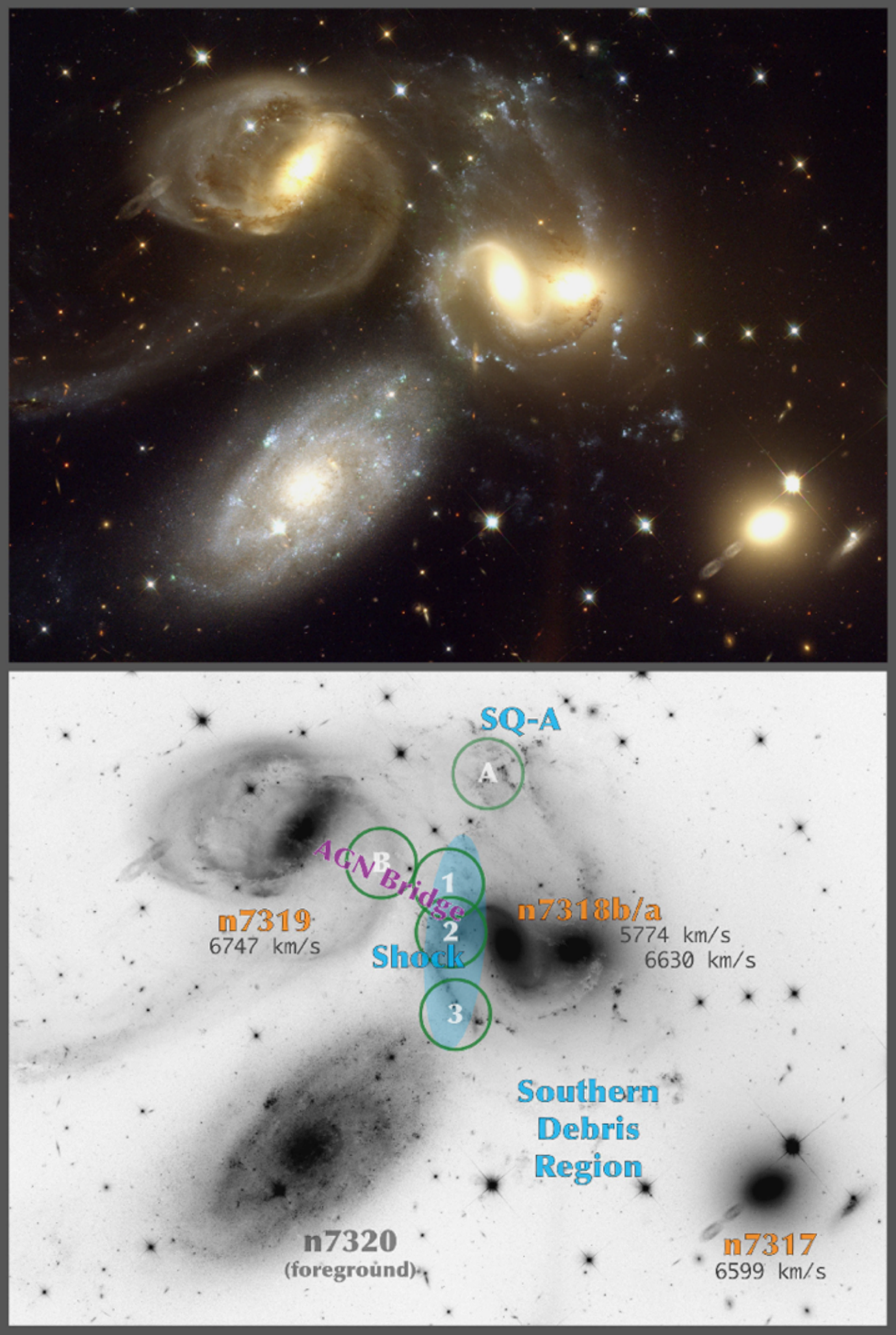}
		\caption{Optical imaging (\bvi-equivalent, \textbf{top}) 
			and annotated grayscale (\textbf{bottom}) map of 
			Stephan's Quintet (HCG~92). North is to the top 
			and east to the left, and the image spans 
			approximately 5\arcmin, or 120~kpc at the adopted 
			distance of 94~Mpc.
			We mark the four galaxies that form the core of 
			the group and their velocities (a fifth galaxy, 
			\n7320C lies to the 
			northeast), a foreground spiral, and four regions 
			of interest: the shock front (pale blue), 
			the southern debris region, X-ray emitter SQ-A, 
			and the `AGN bridge' (see text for information on 
			these regions and related bibliography). Green 
			circles represent the CO pointings of \citet{guillard12}. 
		}\label{fig:finder}
	\end{center}
\end{figure*}

From the above it is clear that the conjunction of these energy input mechanisms is very difficult to interpret in observations and modeling of highly extinguished, nearby star-forming regions. With no clear tipping of the scales in favor of shocks, turbulence, radiation pressure, cooling \citep{bonnell13}, magnetic fields \citep{sellwood99}, and all other mechanisms proposed in the literature, perhaps these potential effects are best studied one at a time. Strong shocks are more common in extra-galactic environments, where friction and collisions can shock-heat large volumes of gas. One of the strongest shock regions in the local universe can be found in the site of the high-speed collision between \n7318B and the intra-group medium (IGM) of Stephan's Quintet (to which we will sometimes refer as ``the Quintet''; see Figure~\ref{fig:finder}), a nearby compact galaxy group (CG; $u_R=6600~$\kms, $d=94~$Mpc if $H_0=70~$\kms\,Mpc$^{-1}$). The high luminosity of the X-ray emission \citep[$L_X = 4\times10^{41}~$erg~s$^{-1}$, ][]{sulentic95} is rather inviting for various investigations not only into the role of shocks as agents of star formation, but also the evolution of galaxy groupings. While the Quintet does not lack the mass required to heat up its IGM, collisions might be the only way to develop an X-ray medium for low-mass compact groups. In this scenario, however, maintaining such an IGM temperature is conditional on consequent collisions and therefore on a continuing association with the interloper \citep[as an upper limit, the gas cools within $\sim1~$Gyr in the simulations of][]{hwang12}. This way a fraction of the IGM can become locked in this hot phase for long and take away a potentially significant part of the reservoir of gas available for star formation. A high-speed collision is then a crucial event in the history of a CG, defining one potential evolutionary pathway that relies on the presence of an IGM \citep[one of the two sequences described by][]{isk10}. If that is the case, then it is worth asking how much gas can get enveloped in such an event, and how does this heat capacity relate to the suppression of star formation on a group-wide scale? 

%======= FIGURE: Finder: Shock region =======

In the Quintet, \citet{appleton06} and \citet{cluver10} mapped the intensity of the warm infrared H$_2$ transition and established this as the dominant cooling pathway in a molecular medium that seems to be forming stars rather inefficiently. This low star formation efficiency (SFE) has also been deduced with UV, optical and IR imaging, with \citet{xu05} reporting a star formation rate $\textup{SFR}=6.7~$\Msun~yr$^{-1}$, of which $30\%$ can be attributed to the bright starburst region known as SQ-A (or the northern starburst region, NSBR), which is thought to have resulted from an interaction between the intruder and \n7318B, taking place before the collision that shocked the IGM \citep{renaud10, hwang12}. However, much of the UV flux on which this SFR measurement is based may be arising from shock excitation lines instead of photoionisation. Context for this SFR is provided by the sub-mm observations by \citet{guillard12} and others \citep{yun97,vm98,leon98,gao_xu_00,smith_struck_01,petitpas05}, who have accounted for the presence of $\approx4\times10^9~$\Msun\ of cool H$_2$ in the shock region. The interpretation on offer is a turbulent cascade process that injects energy within the molecular gas so that its turbulent energy becomes comparable to or higher than its gravitational energy \citep[most of the gas is found to be not only warm, but also diffuse;][]{guillard12, appleton13}. This establishes a causal relation, whereby shocks heat the medium and therefore inhibit the collapse of gas into stars. In this context, the onset, propagation, and energy deposition by shocks is a central piece of the galaxy evolution puzzle, whenever arrangements such as the Quintet come about. 

Conversely, however, if the number of bright, compact, blue sources is a testament to the current state of star formation and its recent history, then the strong shock does not appear to be largely inhibiting the collapse of molecular clouds into stars. The studies of \citet{gallagher01} and \citet{fedotov11} used \hst\ imaging to identify hundreds of star clusters and associations across the shock, SQ-A, and elsewhere, and charted the recent interaction history of this complex system. In terms of shocks, while the X-ray-emitting SQ-A is found to host more star clusters than the shock region, the discrepancy may be due to a longer star formation history, rather than a stronger burst. This was, alas, just beyond the diagnostic reach of past \hst-based studies--although the addition of $U$-band would enable such an investigation. Many results have also been drawn from spectroscopic studies of the Quintet. \citet{trancho12} targeted compact sources in the shock, the star-forming bow and debris region south of \n7318A, as well as the young tidal tail extending to the east of \n7319. In this tail they found a massive, compact cluster of mass $\sim10^6~$\Msun. Their finding of enhanced metallicity for young and currently forming star clusters and associations in the Stephan's Quintet debris system is countered by that of low metallicity gas and stars in the shock region by \citet{iglesias12}. 

These two sets of evidence support contrasting interpretations of the role this shock is playing in either suppressing or advancing the formation of new stars. In this paper we present spectroscopy of some $\approx$\,40~\ha-emitting knots across the Quintet (in the shock region, SQ-A, and the southern debris region), selected from HST-\ha\ images, to help us investigate the link between the large-scale shocks and star formation. We make use of the remarkable wealth of existing broadband coverage and add a much-needed element of spatial resolution. By focussing on spectroscopy we gain access to the velocity axis and can associate these knots with different parts of the complex distribution of the multi-phase gas in the Quintet. Given the broad spatial extent of the hot IGM we examine how any stars can form in this shock region at all, even at a low rate or efficiency. Section~\ref{sec:data} describes the data acquisition and reduction effort. In Section~\ref{sec:measurements} we present our spectral line fitting techniques, including Lick Indices and multi-component fitting of broad-line spectra. We discuss our findings in Sections~\ref{sec:star-formation} and \ref{sec:discussion}, and summarize our interpretation and contribution to the understanding of this well-studied system in Section~\ref{sec:summary}.

\section{Observations and Data Reduction}\label{sec:data}
\begin{figure*}[!t]
	\begin{center}

		\includegraphics[width=\textwidth,angle=0]{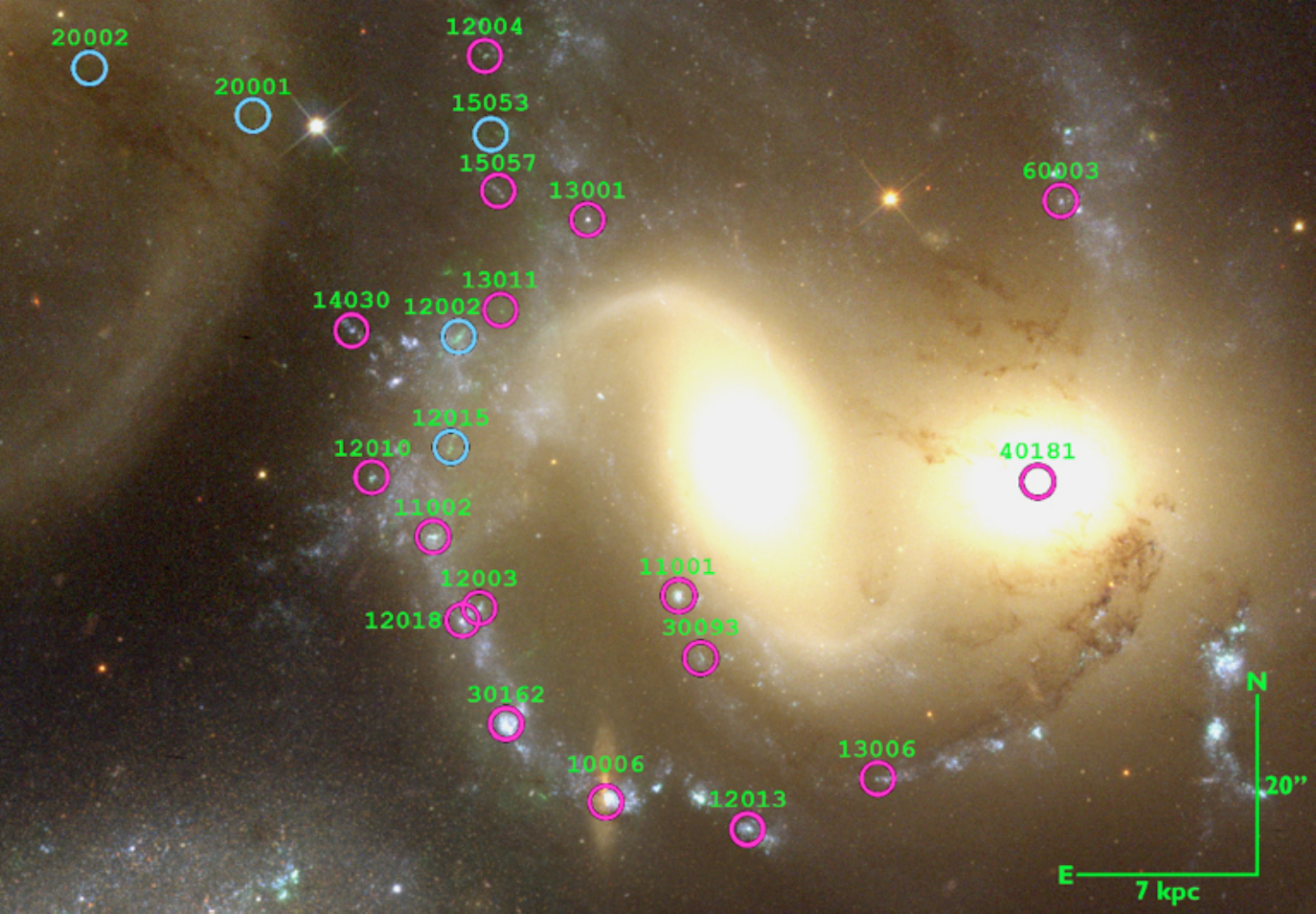}
		\caption{%
			Finding chart showing the approximate placement
			of spectroscopic slits on archival \hst-WFC3 
			imaging. The diameter of the circles corresponds 
			to approximately 4 times the length of the spectroscopic 
			extraction window, which translates to $\approx450~$pc
			at the adopted distance of 94~Mpc. This image, centered 
			on the shock region, is constructed with the F438W, 
			F606W, and F814W images as the blue, green, and 
			red filters. We used narrow-band imaging to select 
			targets (F665N), the resolution of which is not 
			matched by our seeing-limited spectroscopic 
			observations. Blue circles denote broad-line 
			sources (see text).
		}\label{fig:finder_SHK}

	\end{center}
\end{figure*}

\begin{figure*}[!t]
	\begin{center}

		\includegraphics[width=\textwidth,angle=0]{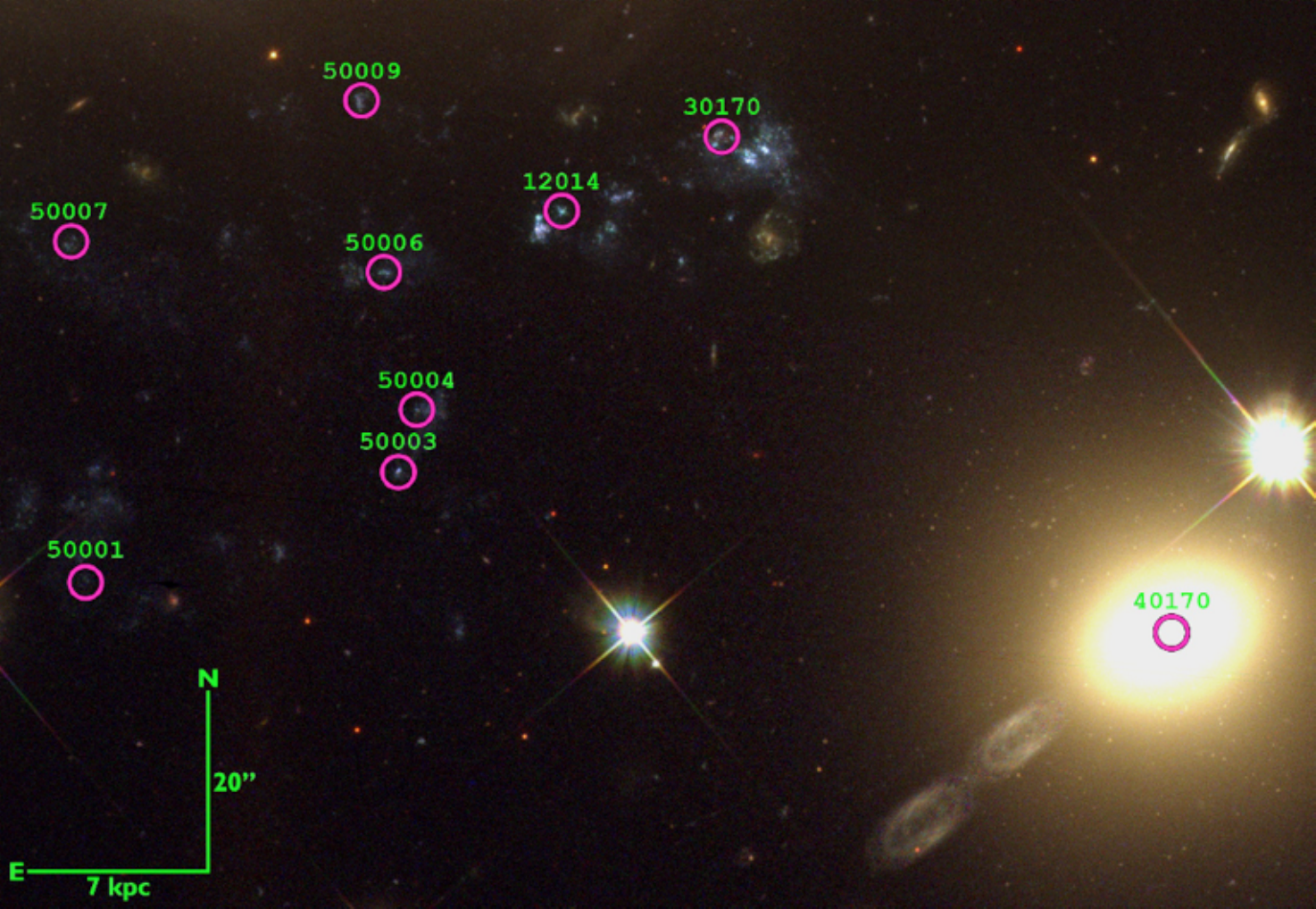}
		\caption{As per Figure~\ref{fig:finder_SHK}, for 
			the Southern Debris Region. These targets 
			represent the faintest in our sample, indicating 
			a less massive gaseous reservoir in this region. 
			The artefact resembling the infinity symbol
			is due to the reflection of the bright star near
			\n7317 (marked as target 40170). 
		}\label{fig:finder_SDR}

	\end{center}
\end{figure*}

\begin{figure*}[!t]
	\begin{center}

		\includegraphics[width=\textwidth,angle=0]{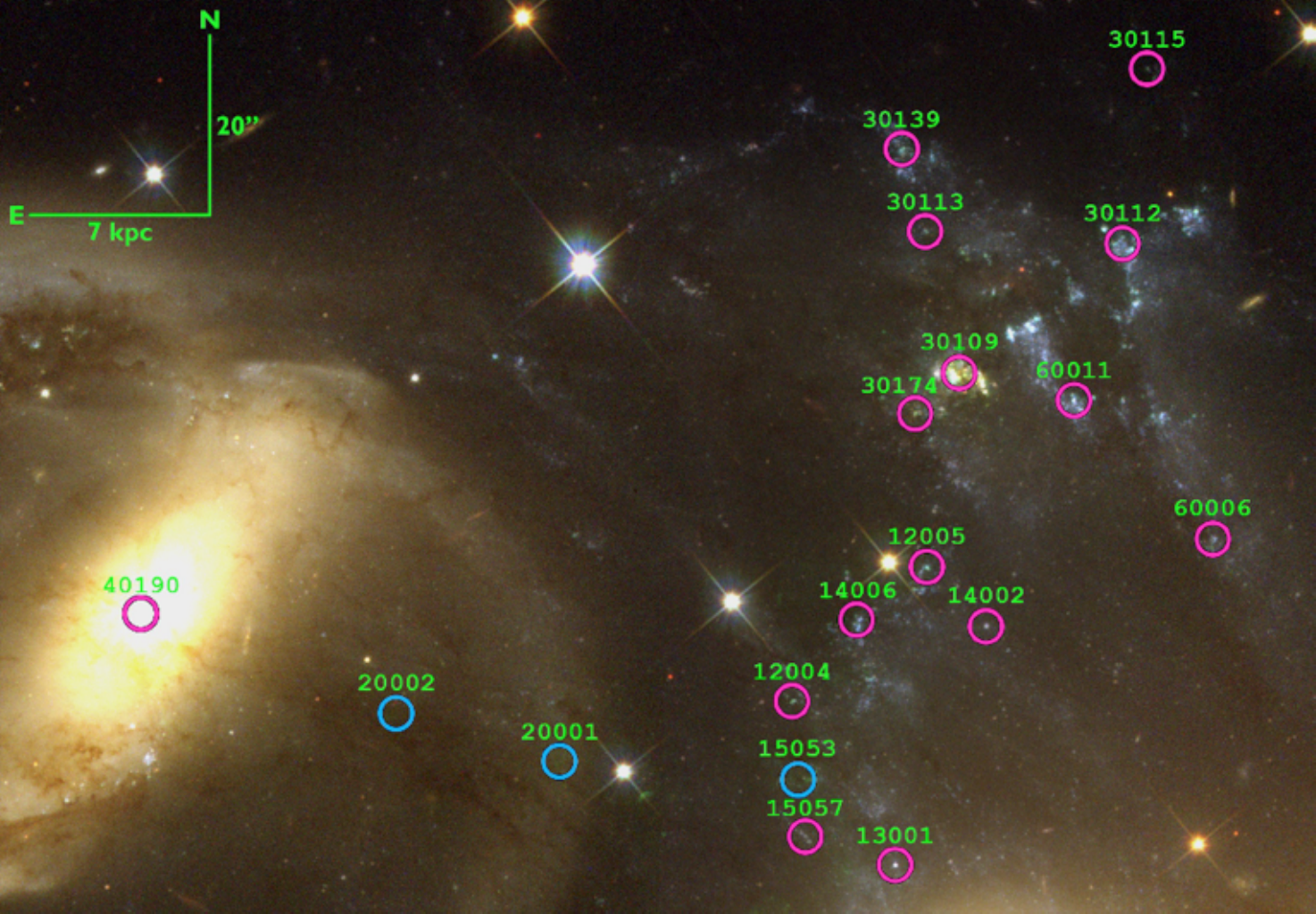}
		\caption{As per Figure~\ref{fig:finder_SHK}, for 
			the northern end of the shock region and 
			SQ-A, the bright starburst region to the North
			of the group core (\n7318A/B). We find much 
			nebulosity in SQ-A, following the relatively 
			high local rate of star formation 
			\citep[1.45~\Msun\,yr$^{-1}$;][]{xu99}. 
			Knots K-30109 and K-30174 have notably yellower 
			colours than the \ha-emitting regions surrounding 
			them, possibly reflecting either higher reddening,
			or a lower metallicity. 
			The velocity of knot K-30113 pushes it out of 
			the transmission of the employed \ha\ filter. 
			This finder also covers the ``AGN bridge'' 
			\citep[][knots K-20001 and K-20002]{cluver10} 
			leading to \n7319 (K-40190). Both ``bridge'' 
			spectra display broad emission lines. 
		}\label{fig:finder_SQA}

	\end{center}
\end{figure*}

%--------------------------------------------------------
This work is based on new spectroscopy taken with Gemini GMOS-N in multi-slit mode as part of program GN-2010B-Q-56 (PI:~\isk), observed in September~2010 under good conditions (typical seeing $0\farcs8$). The spatial coverage is densest in the shock region, however, we sample the rest of the system quite well, with targets in the southern debris region (SDR), SQ-A \citep{xu99}, and the ``AGN bridge'' between the intruder, \n7318B, and \n7319 \citep[a broad-line excitation molecular region connecting the two galaxies in projection;][]{cluver10}. A full journal of observations can be found in Table~\ref{tab:journal}, while Figures~\ref{fig:finder_SHK}, \ref{fig:finder_SDR}, and \ref{fig:finder_SQA} mark all observed sources on color-composite \hst-WFC3 imaging (PID~11502, PI~Noll) focussed on the shock region, SDR, and SQ-A respectively. The RGB filters correspond to F814W, F606W, and F438W. 

%======= TABLE: Journal of Observations =======
\begin{table}[!h]
\caption{Journal of Science Observations for GMOS-N program GN-2010B-Q-56.}
\begin{center}
	\begin{tabular}{lccc}
	\hline
	Date        & Frame ID       & $t_{exp}$ & $\lambda_c$ \\
	            &                &  (s)	      & (\AA) \\
	\hline
	\textit{Mask 1} -- \\
	2010-09-03	& N20100903S0103 & 3130 & 5900\\
	2010-09-03	& N20100903S0104 & 3130 & 5950\\
	2010-09-03	& N20100903S0109 & 3130 & 5900\\
	2010-09-03	& N20100903S0110 & 3130 & 5950\\
	2010-09-06	& N20100906S0092 & 3130 & 5950\\
	\textit{Mask 2} -- \\
	2010-09-03	& N20100903S0115 & 3130 & 5950\\
	2010-09-04	& N20100904S0148 & 3130 & 5900\\
	2010-09-04	& N20100904S0149 & 3130 & 5950\\
	2010-09-04	& N20100904S0154 & 3130 & 5900\\
	2010-09-04	& N20100904S0155 & 3130 & 5950\\
	\textit{Mask 3} -- \\
	2010-09-05	& N20100905S0146 & 3130 & 5950\\
	2010-09-05	& N20100905S0151 & 3130 & 5900\\
	2010-09-05	& N20100905S0152 & 3130 & 5950\\
	2010-09-05	& N20100905S0157 & 3130 & 5900\\
	2010-09-05	& N20100905S0158 & 3130 & 5950\\
	\textit{Mask 4} -- \\
	2010-09-06	& N20100906S0085 & 900	& 5900\\
	2010-09-06	& N20100906S0088 & 900	& 5950\\
	\hline
	\end{tabular}
\end{center}
\label{tab:journal}
\end{table}%

%======= FIGURE: Finder: Shock region =======

%======= FIGURE: Finder: SDR =======

%======= FIGURE: Finder: SQ-A =======

Four multi-slit masks were designed using \texttt{GMMPS}, the Gemini mask making software, to target sources selected manually on the \hst-F665N imaging. The 100~\AA\ bandpass transmits all velocities expected in the system. It reaches $\approx8100~$\kms, albeit the transmission decreases rapidly above $\approx6715~$\kms. We used a brightness criterion for filtering sources, and then accommodated as many as possible on three masks. The fourth mask was designed to target three of the five main galaxies in the Quintet: \n7317, \n7318A, and \n7319, a Seyfert~2 galaxy \citep{veron06}. Slitlets were cut to a $0\farcs75$ width while lengths were customized between $\approx5\arcsec$ and 10\arcsec\ to best accommodate the knots and sufficient background for local sky subtraction. We used the R831 (G5302) grating, which disperses to 0.38 \AA/px ($R\approx4500$). In order to eliminate the inter-chip gaps of the GMOS-N detector, frames were split up into two groups of slightly differing central wavelength: 5900~\AA\ and 5950~\AA. Flat field and arc frames bracketed each science exposure to account for detector flexures and any other variant effects (\eg\ temperature changes) over the course of each individual exposure. Pixels were binned in the spatial direction without loss of resolution, given the pixel size of $\approx0\farcs07$. There was no binning along the spectral axis to preserve resolution. 

\begin{figure*}[tbh]
	\begin{center}

		\includegraphics[width=\textwidth,angle=0]{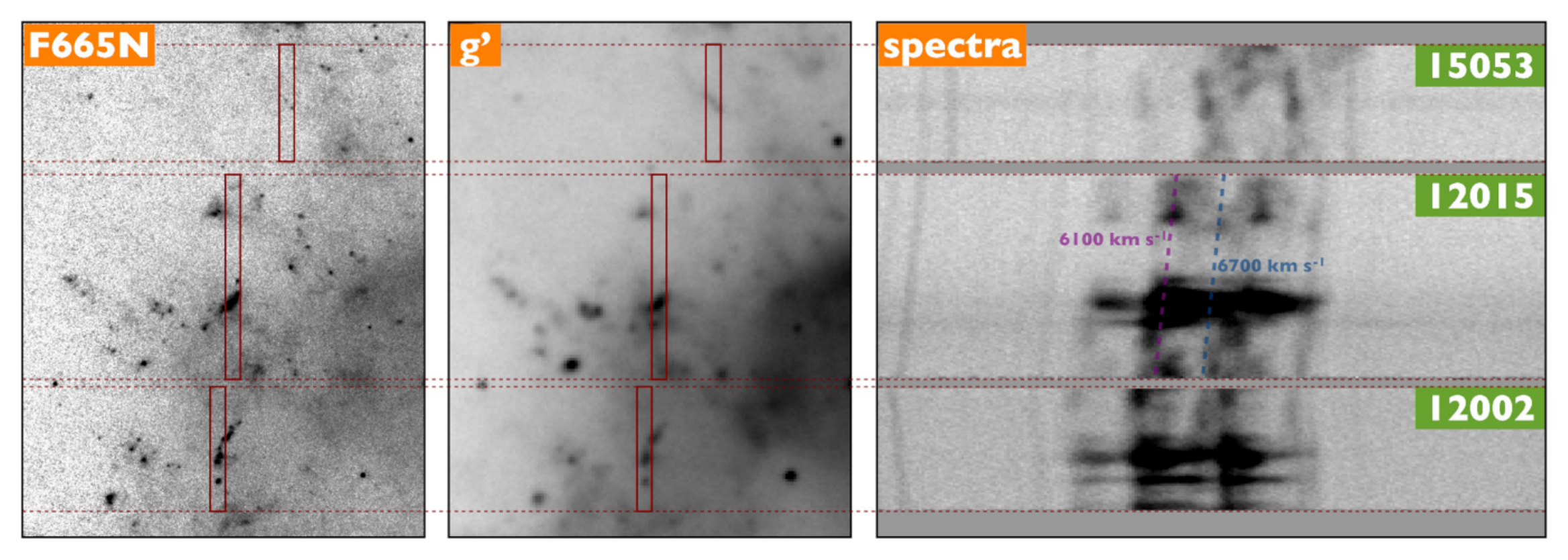}
		\caption{Two-dimensional spectroscopy of the three 
			brightest broad-line emitters, found in three 
			consecutive slits in the shock region. The three
			panels share the same vertical spatial axis, 
			while the third panel replaces the shared x-axis
			with a dispersion (wavelength) axis.  
			On the \textbf{left} we show the \hst-\ha\ image 
			on which the knots were selected, while the 
			\textbf{center} panel shows the GMOS-N $g$-band pre-image, 
			which is more sensitive to diffuse emission 
			features in the background. Such features are 
			seen in the spectra (\textbf{right}; \ha\ bracketed 
			by [N\two]), revealing the 
			presence of a faint, underlying emission component. 
			This will be examined in future work, as in this 
			paper we focus on the knots themselves, and the 
			multiple clumps into which they can be decomposed. 
			As an indication of this substructure we mark two 
			\ha\ features that register at velocities spanning 
			the full range of redshifts seen in the Quintet, 
			apart from  the intruder. 
		}\label{fig:twod}

	\end{center}
\end{figure*}

The spectroscopic data were reduced with a custom pipeline, built around the standard Gemini-IRAF\footnote{
	IRAF is distributed by the National Optical 
	Astronomy Observatories, which are operated 
	by the Association of Universities for Research
	in Astronomy, Inc., under cooperative agreement 
	with the National Science Foundation.} 
reduction process for GMOS-N. After reducing the image frames with \texttt{gsreduce}, offsets were recorded with \texttt{gscut} between the bottom of each image and the lower boundary of the first slit. After this registration process we employed \texttt{mosproc} in two runs: first to perform bias subtraction and a flat field correction; and then to cut individual slits from each image before extracting and tracing each source in a 1\arcsec\ aperture, \ie\ one resolution element (seeing-limited), regardless of the physical size of the source. This corresponds to roughly 450~pc at the adopted distance of 94~Mpc. At the end of this stage we also produced `clean' images of all frames, useful for qualitative, two-dimensional spectroscopy (see Figure~\ref{fig:twod}). The extraction process (\texttt{gsextract}) was first performed on a reference frame, the one that was taken under the best conditions, and the procedure was carbon-copied (with intermediate quality control checks) on all other frames to ensure the application of consistent spectroscopic extraction apertures. The final steps include median-combining all exposures of each spectrum to eliminate chip gaps and cosmic rays, and applying a flux calibration based on a spectroscopic standard star. The above process was then repeated for each mask, before correcting all spectra for foreground extinction from the Milky Way according to the tabulation of \citet{schlegel98}. 

%======= FIGURE: 2D spec =======

The resulting spectral resolution is approximately 40~\kms\ (2.0~\AA) and the spectral coverage typically runs from H$\beta$ to [S\two]. The exact spectral range shifts as a function of offset from the major axis of observation, a line that runs north-south across the Quintet shock region. We record S/N values of $>50$ for emission lines (minimum of 5--10 for the faint [O\one]6300 line), while continuum registers at S/N~$\approx10$. We will outline measurements in the following section, and discuss results in Sections~\ref{sec:star-formation} and \ref{sec:discussion}. The positions of all targets observed and successfully extracted are listed in Table~\ref{tab:combined}, along with measurements and derived information that will be explained in the following sections. 

\begin{deluxetable}{lccccc|cc|c}
\tabletypesize{\scriptsize}
\tablewidth{0pt}
\tablecolumns{10}
\tablecaption{%
  Positional information and derived properties of all observed targets.%
  \label{tab:combined}%
}
\tablehead{
  \colhead{ID} &    
  \colhead{$\alpha, \delta$} & 
  \colhead{Type} & 
  \colhead{$u({\textup{\scriptsize H}\alpha})$} &
  \colhead{$\sigma({\textup{\scriptsize H}\alpha})$} & 
  \colhead{$A_V$} & 
  \colhead{SFR} & 
  \colhead{$M^*$} & 
  %\colhead{$N_\textup{\scriptsize p}$} & 
  \colhead{$L_{\textup{\scriptsize H}\alpha}$} \\
  &
  \colhead{(h~m~s)~~($^\odot~\arcmin~\arcsec$)} &
  &
  \tmult{(\kms)} &
  \colhead{(mag)} &
  \colhead{($\times10^{-3}$~\Msun~yr$^{-1}$)} &
  \colhead{($\times10^{6}~\Msun$)} & 
  %\colhead{($\times10^{50}$)} & 
  \colhead{$(L_{30\,\textup{Dor}})$}
}
\startdata
10006 & 22~35~59.196~~+33~57~33.98 & E &                    5668  &   49    & 0.66   &    27.5 &    2.9  & 0.65 \\%     &   15.83 
%11001& 22~35~58.795~~+33~57~48.29 & E &                    5698  &   17    & 1.58   & \nodata &    1.4  & 0.22 \\%    &    5.31 
11002 & 22~36~00.185~~+33~57~52.52 & E &                    5764  &   48    & 0.98   &     5.3 &    2.8  & 0.10 \\%     &    2.36 
12002 & 22~36~00.043~~+33~58~06.74 & \phantom{$^*$}E$^*$ &\nodata & \nodata &\nodata & \nodata & \nodata & \nodata \\%   & \nodata 
12004 & 22~35~59.890~~+33~58~26.76 & E &                    6020  &   44    & 1.04   &     2.3 &    2.0  & 0.04 \\%     &    0.99 
12005 & 22~35~59.126~~+33~58~36.24 & E &                    6049  &   44    & 0.70   &     2.0 &    3.3  & 0.05 \\%     &    1.11 
12010 & 22~36~00.550~~+33~57~56.66 & E &                    5838  &   44    & 0.72   &     2.0 &    2.2  & 0.05 \\%     &    1.10 
12013 & 22~35~58.392~~+33~57~31.73 & E &                    5647  &   46    & 0.70   &     1.9 &    2.5  & 0.05 \\%     &    1.11 
12014 & 22~35~55.346~~+33~57~11.79 & E &                    5766  &   49    & 0.32   &     1.5 &    6.1  & 0.05 \\%     &    1.09 
12015 & 22~36~00.094~~+33~57~58.82 & \phantom{$^*$}E$^*$ &\nodata & \nodata &\nodata & \nodata & \nodata & \nodata \\%   & \nodata 
12018 & 22~36~00.029~~+33~57~46.49 & A/E &                \nodata & \nodata &\nodata & \nodata & \nodata & \nodata \\%   & \nodata 
13001 & 22~35~59.309~~+33~58~15.08 & A/E &                \nodata & \nodata &\nodata & \nodata & \nodata & \nodata \\%   & \nodata 
13003 & 22~35~59.930~~+33~57~47.36 & E &                    5762  &   48    & 0.23   &     1.1 &    5.6  & 0.04 \\%     &    0.91 
13006 & 22~35~57.658~~+33~57~35.22 & E &                    5610  &   55    & 1.08   &     0.5 & \nodata & 0.01 \\%     &    0.21 
13011 & 22~35~59.803~~+33~58~08.53 & E &                    6665  &   89    &\nodata & \nodata & \nodata & \nodata \\%   & \nodata 
14002 & 22~35~58.788~~+33~58~32.09 & A/E &                \nodata & \nodata &\nodata & \nodata & \nodata & \nodata \\%   & \nodata 
14006 & 22~35~59.532~~+33~58~32.80 & E &                    6023  &   47    & 0.82   &     1.0 &    2.3  & 0.02 \\%     &    0.50 
14030 & 22~36~00.653~~+33~58~07.16 & A/E &                \nodata & \nodata &\nodata & \nodata & \nodata & \nodata \\%   & \nodata 
15053 & 22~35~59.856~~+33~58~20.69 & \phantom{$^*$}E$^*$ &\nodata & \nodata &\nodata & \nodata & \nodata & \nodata \\%   & \nodata 
15057 & 22~35~59.798~~+33~58~16.73 & A/E &                \nodata & \nodata &\nodata & \nodata & \nodata & \nodata \\%   & \nodata 
20001 & 22~36~01.219~~+33~58~22.61 & \phantom{$^*$}E$^*$ &\nodata & \nodata &\nodata & \nodata & \nodata & \nodata \\%   & \nodata 
20002 & 22~36~02.150~~+33~58~25.64 & \phantom{$^*$}E$^*$ &\nodata & \nodata &\nodata & \nodata & \nodata & \nodata \\%   & \nodata 
30093 & 22~35~58.658~~+33~57~43.95 & A/E &                \nodata & \nodata &\nodata & \nodata & \nodata & \nodata \\%   & \nodata 
30109 & 22~35~58.910~~+33~58~49.92 & E &                    6659  &   59    & 1.25   &    20.2 &    2.1  & 0.33 \\%     &    8.06 
30112 & 22~35~58.008~~+33~58~59.71 & E &                    5989  &   44    & 0.40   &     2.4 &    4.7  & 0.07 \\%     &    1.67 
30113 & 22~35~59.138~~+33~59~00.20 & E &                    6858  &   42    & 1.94   &     1.3 &    1.3  & 0.01 \\%     &    0.30 
30115 & 22~35~57.862~~+33~59~11.68 & E &                    5988  &   51    & 0.72   &     0.5 &    3.9  & 0.01 \\%     &    0.26 
30116 & 22~36~07.567~~+33~59~21.37 & A/E &                \nodata & \nodata &\nodata & \nodata & \nodata & \nodata \\%   & \nodata 
30139 & 22~35~59.256~~+33~59~06.02 & E &                    6030  &   47    & 0.72   &     1.9 &    4.1  & 0.04 \\%     &    1.04 
30162 & 22~35~59.798~~+33~57~39.35 & E &                    5710  &   47    & 0.95   &     1.0 &    2.2  & 0.02 \\%     &    0.57 
30170 & 22~35~54.384~~+33~57~17.02 & E &                    5778  &   45    &\nodata & \nodata &         & \nodata \\%   & \nodata 
30174 & 22~35~59.194~~+33~58~47.04 & E &                    6649  &   45    & 2.26   &     7.1 &    0.8  & 0.05 \\%     &    1.26 
%40170& 22~35~51.881~~+33~56~41.74 & A &                  \nodata & \nodata &\nodata & \nodata & \nodata & \nodata \\%    \nodata &
%40181& 22~35~56.743~~+33~57~56.22 & A &                    5687  & \fbo}   &\nodata & \nodata & \nodata & 0.09 \\%       2.16   &
%40190& 22~36~03.588~~+33~58~33.02 & \phantom{$^*$}E$^*$ &\nodata & \nodata &\nodata & \nodata & \nodata & \nodata \\%    \nodata &
50001 & 22~35~58.068~~+33~56~45.48 & E &                    5687  &   42    & 1.44   &     0.4 & \nodata & 0.01 \\%     &    0.15 
50003 & 22~35~56.282~~+33~56~53.37 & E &                    5634  &   55    & 3.26   &     1.3 & \nodata & 0.01 \\%     &    0.11 
50006 & 22~35~56.366~~+33~57~07.48 & E &                    5670  &   47    & 0.59   &     0.6 &    4.8  & 0.02 \\%     &    0.39 
50007 & 22~35~58.157~~+33~57~09.65 & E &                    5645  &   44    &\nodata &     0.1 & \nodata & 0.01 \\%     &    0.11 
50009 & 22~35~56.501~~+33~57~19.57 & E &                    5695  &   52    & 2.51   &     1.1 & \nodata & 0.01 \\%     &    0.18 
60003 & 22~35~56.597~~+33~58~16.35 & A/E &                \nodata & \nodata &\nodata & \nodata & \nodata & \nodata \\%   & \nodata 
60006 & 22~35~57.485~~+33~58~38.16 & E &                    5998  &   45    & 1.35   &     1.5 &    2.2  & 0.02 \\%     &    0.50 
60011 & 22~35~58.284~~+33~58~48.17 & E &                    6000  &   46    &\nodata &     0.8 & \nodata & 0.03 %       &    0.72 
\enddata
\vspace{-10pt}\tablecomments{`E' and `A' in the `Type' 
  column stand for emission and absorption, and `*' denotes 
  broad-line emission, defined as being broader than 
  the instrumental resolution of $\approx2~$\AA, or $45~$\kms
  at \ha. Velocity dispersions are deconvolved with the 
  instrumental signature.
  The faint knot~13011 may be a broad-line emitter. 
  `A/E' sources do not allow for spectral fitting, while 
  the kinematical structure of E$^*$ clumps is listed in 
  Table~\ref{tab:broadline}. 
  $A_V$ measurements were not possible for sources with 
  low signal in the H$\beta$ region. 
  The columns following the first vertical separator only 
  apply to H\two\ regions. 
  \ha-to-SFR conversions are based on \citet{kennicutt98araa}. 
  The final column lists the \ha\ luminosity normalised to nearby
  starburst region 30~Dor. 
  }
\end{deluxetable}

\begin{figure*}[!t]
	\begin{center}

		\includegraphics[width=0.9\textwidth,angle=0]{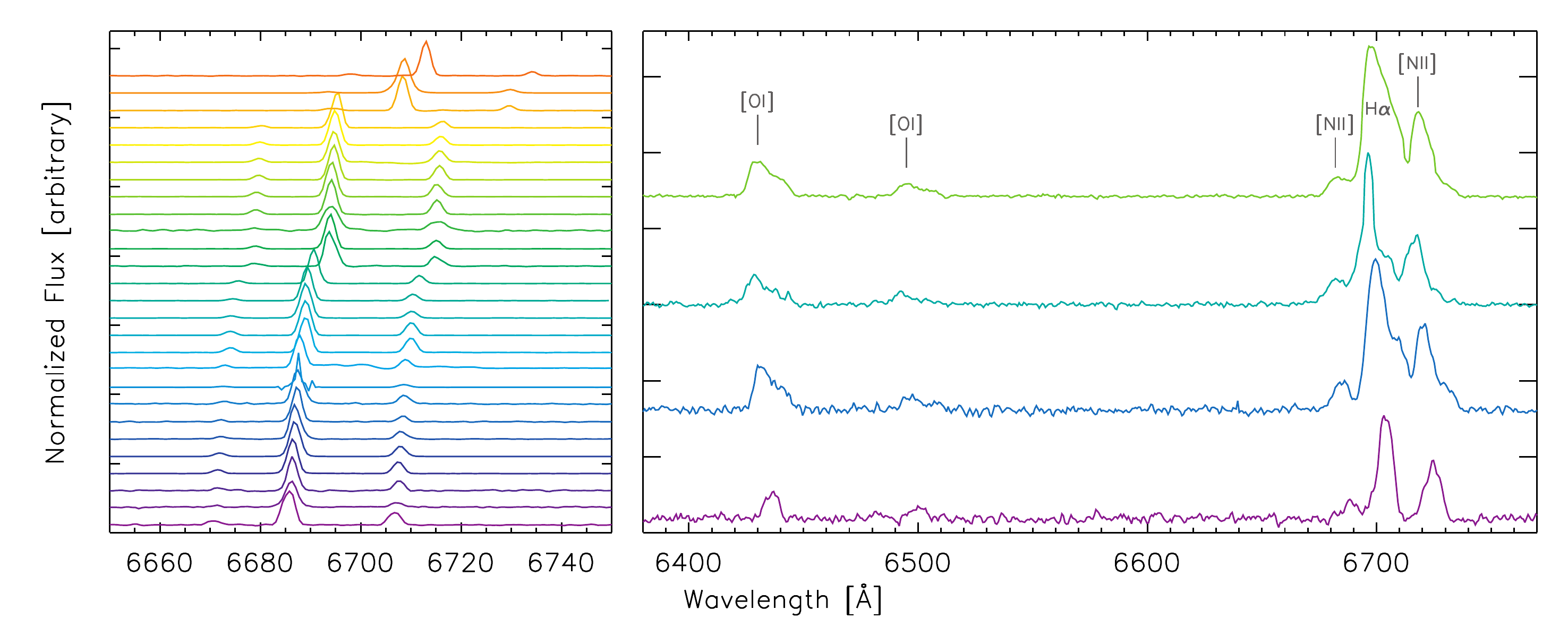}
		\caption{Normalized spectra of H\two\ regions 
			(\textbf{left}) and broad-line knots 
			(\textbf{right}) in Stephan's Quintet, with 
			differently stretched wavelength axes. We exclude 
			eight featureless spectra of stellar associations 
			currently emerging from their cocoons of natal 
			gas. The plots are focussed on the spectral 
			region about the \ha\ and [N\two] lines, plus 
			the [O\one]6300,6340 lines for broad-line spectra. 
			The H\two\ regions are ordered in 
			increasing redshift to reveal various layers 
			of star-forming material along our line of 
			sight. A shallow velocity gradient is evident 
			in the bottom half of this plot, representing 
			knots in the SDR and the lower arc of the shock 
			region, which coincide with the velocity of the 
			intruder, \n7318B. The upper half of the left 
			panel is populated mostly by knots in SQ-A, 
			including the top spectrum, at a redshift 
			accordant with a recently discovered CO feature 
			\citep{guillard12}. The two spectra below the 
			top are in the systemic velocity of the IGM. 
			Narrow-line emitters will be studied in 
			Section~\ref{sec:star-formation}. 
			The broad-line spectra are presented in order 
			of increasing flux (bottom to top). These broad, 
			irregular spectral profiles will be examined 
			in Section~\ref{sec:pan} and Figure~\ref{fig:pan}. 
		}\label{fig:spectra}

	\end{center}
\end{figure*}

\begin{figure*}[!t]
	\begin{center}

		\includegraphics[width=0.9\textwidth,angle=0]{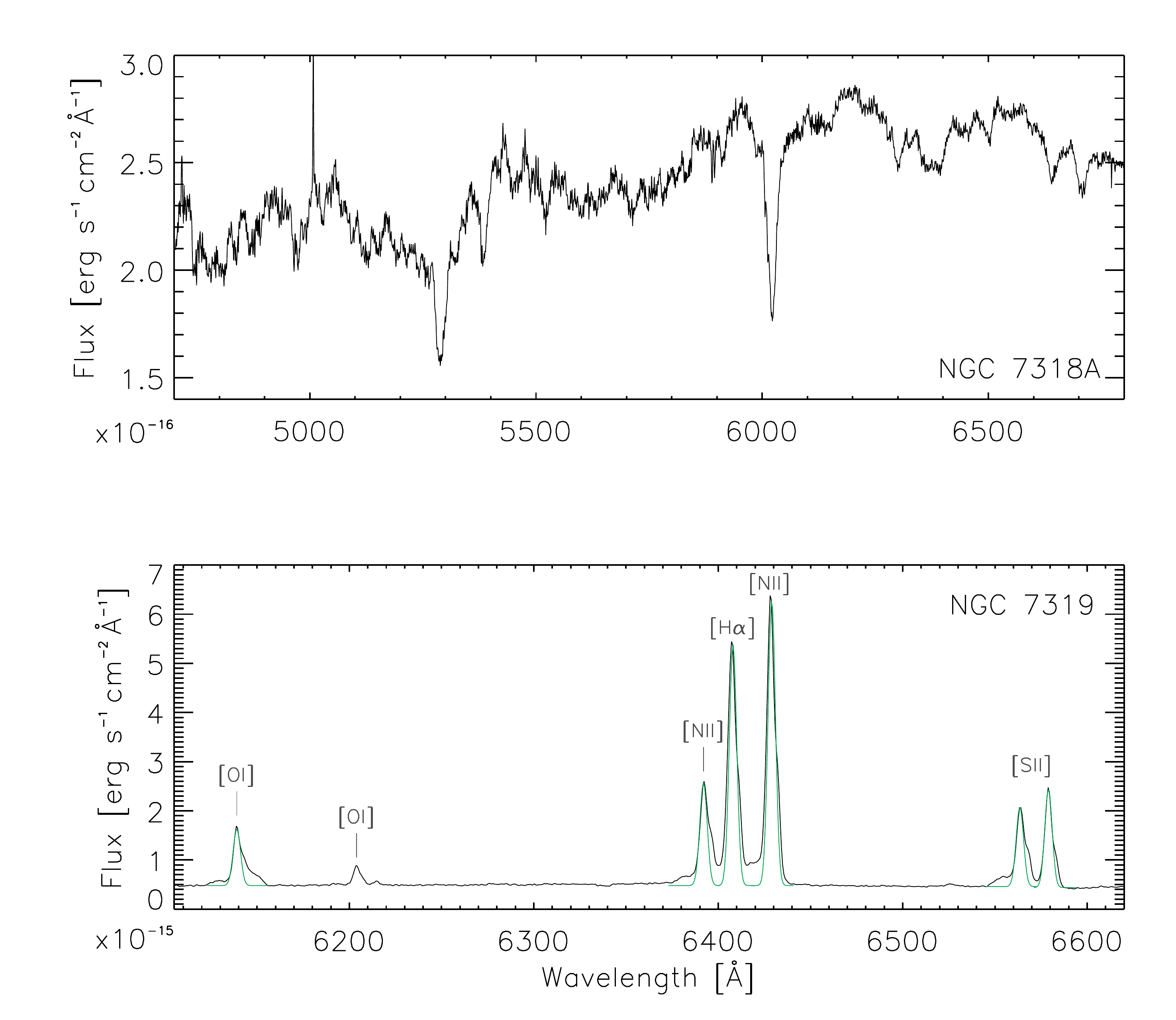}
		\caption{Flux-calibrated spectra of \n7318A (\textbf{top})
			and \n7319 (\textbf{bottom}). There is no trace of 
			emission-line structure in \n7318A, branding this 
			galaxy as early type. The 5007~\AA\ sky line has not 
			been entirely subtracted due to the small available 
			sky window in the slit (the observing program was 
			designed for compact sources). \n7319 shows no
			line broadening, consistent with its Seyfert~2 
			classification, but does feature blue and red shoulders
			on nebular lines, most prominently [O\one]6400, and 
			the bluer halves of the [N\two] and [S\two] doublets. 
			We have drawn rough Gaussian fits representing the 
			instrumental resolution at the marked lines.
		}\label{fig:7318a}

	\end{center}
\end{figure*}

Figure~\ref{fig:spectra} shows the \ha\ region of all spectra of compact sources. Most (27) appear H\two\ region-like, with the exception of five knots that exhibit emission lines broadened to the point of blending with neighboring spectral features. Four broad-line emitters are shown here, as the fifth was not bright enough for a successful extraction. The occasionally complex velocity-space structure is evident in right panel of Figure~\ref{fig:twod}. Additionally, we show the flux-calibrated spectrum of \n7318A in Figure~\ref{fig:7318a}, the first to be published of this system. The lack of emission lines confirms it as an early-type galaxy, as morphologically classified by \citet{sulentic01}. This indicates that the dust features seen across the face of this galaxy are in fact part of \n7318B, back-lit by \n7318A \citep[see][for examples of this process]{keel13}. The other two galaxies covered show no surprising traits: the Seyfert signature is evident in the bright lines emerging from the nuclear region of \n7319 (also shown on Figure~\ref{fig:7318a}), along with blue and red `shoulders' on the nebular lines indicative of out-flowing material \citep[as noted by][]{aoki98}. \n7317 shows no features unexpected from its classification as an elliptical galaxy. 

%======= FIGURE: 1D spec =======

%======= FIGURE: NGC 7318A spectrum =======

%-----------------------------------------------------
\section{Sample Definition and Spectral Line Fitting}\label{sec:measurements}
%-----------------------------------------------------

\subsection{Sample Completeness\\}\label{sec:completeness}
%-----------------------------------
Since our spectroscopic slits sample discrete regions across the Quintet, we use the catalogue of \citet{fedotov11}, which reaches fainter sources with imaging, to estimate the completeness of our sample. More specifically, we make use of new SED fitting of \hst-based UBVI photometry, an extension to their star cluster catalogue (Fedotov~\etal, in preparation), which includes derivations of age and mass for each source. Given the $20\times$ poorer resolution of our ground-based spectroscopy, we draw the slit extraction windows on HST-images and simply count the number of \citeauthor{fedotov11} sources that fall within these windows. Since \ha\ emission was a prerequisite for the selection of our sample, we consider only those sources in the \citeauthor{fedotov11} catalogue younger than 10~Myr. 

We estimate that our \ha-emitters represent approximately 9.6\% and 13.6\% by number of all H\two\ regions in the shock region and SQ-A respectively. Elsewhere the coverage is low ($\approx\,$2.5\% overall), owing to the sparsity of the SDR and other regions, and the more dense coverage of the shock region and SQ-A in our slit masks. Due to the nature of the observations (spectroscopy versus imaging), our sample is biased toward the most luminous (massive) H\two\ regions. A more extensive, quantitative comparison of the two samples will be presented in Fedotov \etal\ (in preparation). 

\subsection{Measurement of Radial Velocity, Extinction, and Line Indices}\label{sec:hiifit}
%-----------------------------------
We measured radial velocities of emission-line sources with the \texttt{EMSAO} task in the \texttt{IRAF-RVSAO} package, and took the H$\alpha$ velocity to represent each knot (following the H$\alpha$-based selection). We then measured the Balmer decrement to derive the reddening of each target through a nominal H$\alpha$/H$\beta$ flux ratio of 2.85, and using the extinction laws of \citet{seaton79} and \citet{howarth83}. Kinematical measurements can be found in Table~\ref{tab:combined}. The `A/E' type sources in our sample display a noisy continuum, rather than clear emission or absorption features, hence no measurements were possible. The majority of these appear shaped by a stellar continuum with absorption lines filled, which leads us to assume they are young stellar associations emerging from their natal shrouds (\ie, the remainder of the gas from which the stars formed). In the case of \n7318A we employed the \texttt{RVSAO-XCSAO} task to cross-correlate its absorption-line spectrum (Figure~\ref{fig:7318a}) with four stellar templates from the default library. The velocity was derived as the median of the four values, $6622.5\pm35.9~$km/s. This uncertainty is computed as the square addition of the individual errors of approximately $18~$\kms. This is consistent with the $6630\pm23~$\kms\ literature value of \citet{hickson92}. 

%======= TABLE: IndexF =======
\begin{deluxetable}{l r@{$\pm$}l r@{$\pm$}l r@{$\pm$}l r@{$\pm$}l r@{$\pm$}l r@{$\pm$}l r@{$\pm$}l r@{$\pm$}l r@{$\pm$}l}
\rotate
\tabletypesize{\tiny}
\tablewidth{0pt}
\tablecolumns{19}
\tablecaption{Emission Line Fits with Lick Indices.\label{tab:indexf}}

\tablehead{
  \colhead{ID} &
  \tmult{H$\beta$} & 
  \tmult{O\three}  & 
  \tmult{O\three}  & 
  \tmult{O\one}    & 
  \tmult{N\two}    & 
  \tmult{\ha}      & 
  \tmult{N\two}    & 
  \tmult{S\two}    & 
  \tmult{S\two}    \\
  & 
  \tmult{$\lambda4860$} &
  \tmult{$\lambda4959$} &
  \tmult{$\lambda5007$} & 
  \tmult{$\lambda6300$} &
  \tmult{$\lambda6548$} & 
  \tmult{$\lambda6563$} & 
  \tmult{$\lambda6583$} & 
  \tmult{$\lambda6716$} & 
  \tmult{$\lambda6731$}
  }

\startdata
10006  &  0.5285  &  0.0011  &  0.2676  &  0.0007  &  0.7977  &  0.0019  &  0.0207  &  0.0013  &  0.2010  &  0.0052  &  2.0402  &  0.0167  &  0.5872  &  0.0006  &  0.1804  &  0.0084  &  0.1414  &  0.0035    \\
11001  &  0.1161  &  0.0024  &  0.0253  &  0.0010  &  0.0777  &  0.0011  &  0.0010  &  0.0005  &  0.0414  &  0.0034  &  0.6849  &  0.0034  &  0.1265  &  0.0008  &  0.0327  &  0.0015  &  0.0234  &  0.0009    \\
11002  &  0.0680  &  0.0010  &  0.0174  &  0.0003  &  0.0539  &  0.0002  &  0.0032  &  0.0002  &  0.0345  &  0.0008  &  0.3046  &  0.0055  &  0.1070  &  0.0003  &  0.0413  &  0.0021  &  0.0281  &  0.0019    \\
12002  &  0.0370  &  0.0012  &  0.0087  &  0.0005  &  0.0298  &  0.0010  &  0.0519  &  0.0043  &  0.0359  &  0.0157  &  0.2553  &  0.0180  &  0.1523  &  0.0192  &  0.1221  &  0.0162  &  0.0734  &  0.0122    \\
12004  &  0.0277  &  0.0008  &  0.0089  &  0.0004  &  0.0131  &  0.0011  &  0.0009  &  0.0002  &  0.0170  &  0.0003  &  0.1272  &  0.0027  &  0.0497  &  0.0002  &  0.0175  &  0.0008  &  0.0124  &  0.0003    \\
12005  &  0.0364  &  0.0008  &  0.0316  &  0.0005  &  0.1031  &  0.0001  &  0.0025  &  0.0002  &  0.0093  &  0.0003  &  0.1429  &  0.0009  &  0.0272  &  0.0002  &  0.0145  &  0.0009  &  0.0103  &  0.0004    \\
12010  &  0.0356  &  0.0004  &  0.0178  &  0.0004  &  0.0549  &  0.0004  &  0.0011  &  0.0001  &  0.0109  &  0.0002  &  0.1414  &  0.0016  &  0.0323  &  0.0001  &  0.0160  &  0.0005  &  0.0113  &  0.0003    \\
12013  &  0.0364  &  0.0010  &  0.0135  &  0.0004  &  0.0337  &  0.0006  &  0.0013  &  0.0001  &  0.0157  &  0.0004  &  0.1430  &  0.0024  &  0.0480  &  0.0006  &  0.0176  &  0.0012  &  0.0133  &  0.0009    \\
12014  &  0.0428  &  0.0004  &  0.0277  &  0.0004  &  0.0823  &  0.0006  &  0.0040  &  0.0001  &  0.0090  &  0.0004  &  0.1411  &  0.0015  &  0.0273  &  0.0001  &  \tmult{\nodata}    &  \tmult{\nodata}      \\
12015  &  0.0046  &  0.0007  &  0.0050  &  0.0011  &  0.0041  &  0.0006  &  0.0177  &  0.0014  &  0.0166  &  0.0036  &  0.0806  &  0.0030  &  0.0511  &  0.0039  &  0.0373  &  0.0037  &  0.0253  &  0.0026    \\
12018  &  \tmult{\nodata}    &  -0.0020 &  0.0003  &  0.0013  &  0.0010  &  -0.0032 &  0.0004  &  -0.0001 &  0.0008  &  0.0015  &  0.0005  &  0.0013  &  0.0002  &  -0.0005 &  0.0006  &  -0.0001 &  0.0003    \\
13001  &  -0.0004 &  0.0015  &  -0.0085 &  0.0003  &  -0.0004 &  0.0004  &  -0.0006 &  0.0004  &  -0.0002 &  0.0001  &  -0.0003 &  0.0002  &  -0.0001 &  0.0001  &  0.0011  &  0.0011  &  -0.0012 &  0.0006    \\
13003  &  0.0370  &  0.0008  &  0.0086  &  0.0003  &  0.0209  &  0.0004  &  0.0016  &  0.0002  &  0.0156  &  0.0003  &  0.1171  &  0.0016  &  0.0490  &  0.0001  &  0.0216  &  0.0006  &  0.0154  &  0.0005    \\
13006  &  0.0059  &  0.0005  &  0.0046  &  0.0003  &  0.0115  &  0.0006  &  0.0002  &  0.0002  &  0.0035  &  0.0002  &  0.0277  &  0.0005  &  0.0099  &  0.0001  &  \tmult{\nodata}    &  \tmult{\nodata}      \\
13011  &  0.0076  &  0.0004  &  -0.0026 &  0.0002  &  0.0029  &  0.0005  &  0.0012  &  0.0001  &  0.0077  &  0.0022  &  0.0178  &  0.0017  &  0.0032  &  0.0006  &  0.0045  &  0.0007  &  0.0007  &  0.0004    \\ 
14002  &  -0.0046 &  0.0029  &  0.0027  &  0.0009  &  0.0044  &  0.0004  &  -0.0006 &  0.0002  &  -0.0000 &  0.0004  &  0.0013  &  0.0004  &  0.0022  &  0.0002  &  -0.0006 &  0.0002  &  0.0000  &  0.0002    \\
14006  &  0.0154  &  0.0012  &  0.0038  &  0.0002  &  0.0215  &  0.0004  &  0.0016  &  0.0002  &  0.0083  &  0.0002  &  0.0639  &  0.0016  &  0.0193  &  0.0002  &  0.0124  &  0.0002  &  0.0060  &  0.0003    \\
14030  &  0.0068  &  0.0005  &  0.0178  &  0.0005  &  0.0165  &  0.0006  &  0.0015  &  0.0001  &  0.0019  &  0.0001  &  0.0172  &  0.0003  &  0.0073  &  0.0003  &  0.0052  &  0.0002  &  0.0036  &  0.0002    \\
15053  &  0.0248  &  0.0012  &  0.0065  &  0.0004  &  0.0156  &  0.0010  &  0.0194  &  0.0022  &  0.0130  &  0.0043  &  0.0774  &  0.0058  &  0.0518  &  0.0046  &  0.0350  &  0.0047  &  0.0226  &  0.0022    \\
15057  &  \tmult{\nodata}    &  0.0039  &  0.0005  &  0.0026  &  0.0004  &  0.0004  &  0.0003  &  0.0000  &  0.0001  &  0.0007  &  0.0001  &  -0.0002 &  0.0002  &  -0.0016 &  0.0012  &  -0.0021 &  0.0006    \\
20001  &  0.0127  &  0.0007  &  -0.0021 &  0.0004  &  0.0015  &  0.0006  &  0.0060  &  0.0002  &  0.0058  &  0.0020  &  0.0276  &  0.0007  &  0.0149  &  0.0004  &  0.0129  &  0.0006  &  0.0056  &  0.0007    \\
20002  &  \tmult{\nodata}    &  \tmult{\nodata}    &  \tmult{\nodata}    &  0.0007  &  0.0003  &  -0.0001 &  0.0002  &  -0.0002 &  0.0003  &  0.0009  &  0.0002  &  0.0055  &  0.0013  &  0.0102  &  0.0024    \\
30093  &  -0.0675 &  0.0318  &  -0.0478 &  0.0457  &  -0.0655 &  0.0191  &  0.0396  &  0.0077  &  -0.0139 &  0.0045  &  0.0059  &  0.0025  &  -0.0058 &  0.0060  &  -0.0269 &  0.0164  &  -0.0224 &  0.0129    \\
30109  &  0.2050  &  0.0009  &  0.1875  &  0.0009  &  0.6235  &  0.0008  &  0.0107  &  0.0004  &  0.0532  &  0.0457  &  1.0383  &  0.0032  &  0.1117  &  0.0023  &  0.0611  &  0.0042  &  0.0382  &  0.0044    \\
30112  &  0.0631  &  0.0012  &  0.0417  &  0.0015  &  0.1244  &  0.0005  &  0.0018  &  0.0003  &  0.0175  &  0.0003  &  0.2158  &  0.0023  &  0.0546  &  0.0002  &  \tmult{\nodata}    &  \tmult{\nodata}      \\
30113  &  0.0055  &  0.0014  &  0.0153  &  0.0003  &  0.0347  &  0.0004  &  0.0007  &  0.0002  &  0.0023  &  0.0002  &  0.0383  &  0.0002  &  0.0044  &  0.0002  &  0.0015  &  0.0003  &  0.0018  &  0.0001    \\
30115  &  0.0084  &  0.0004  &  0.0066  &  0.0004  &  0.0122  &  0.0004  &  0.0002  &  0.0002  &  0.0028  &  0.0002  &  0.0333  &  0.0004  &  0.0094  &  0.0001  &  \tmult{\nodata}    &  \tmult{\nodata}      \\
30116  &  \tmult{\nodata}    &  \tmult{\nodata}    &  \tmult{\nodata}    &  0.0010  &  0.0004  &  0.0003  &  0.0001  &  -0.0002 &  0.0001  &  0.0000  &  0.0003  &  -0.0000 &  0.0001  &  0.0002  &  0.0002    \\
30139  &  0.0339  &  0.0016  &  0.0251  &  0.0011  &  0.0662  &  0.0005  &  0.0031  &  0.0002  &  0.0114  &  0.0018  &  0.1344  &  0.0014  &  0.0345  &  0.0002  &  0.0206  &  0.0018  &  0.0141  &  0.0010    \\
30162  &  0.0167  &  0.0006  &  0.0072  &  0.0004  &  0.0216  &  0.0006  &  0.0008  &  0.0010  &  0.0056  &  0.0011  &  0.0738  &  0.0040  &  0.0317  &  0.0025  &  0.0086  &  0.0020  &  0.0096  &  0.0013    \\
30170  &  0.0408  &  0.0015  &  0.0395  &  0.0002  &  0.1088  &  0.0002  &  0.0034  &  0.0006  &  \tmult{\nodata}    &  \tmult{\nodata}    &  \tmult{\nodata}    &  \tmult{\nodata}    &  \tmult{\nodata}      \\
30174  &  0.0202  &  0.0010  &  0.0221  &  0.0002  &  0.0683  &  0.0003  &  0.0020  &  0.0003  &  0.0156  &  0.0018  &  0.1629  &  0.0024  &  0.0235  &  0.0009  &  0.0133  &  0.0010  &  0.0087  &  0.0005    \\
%40170  &  0.0221  &  0.0091  &  -0.3913 &  0.0259  &  0.0342  &  0.0188  &  -0.1735 &  0.0073  &  -0.0364 &  0.0102  &  -0.0781 &  0.0110  &  -0.0605 &  0.0019  &  \tmult{\nodata}    &  \tmult{\nodata}      \\
%40181  &  0.0867  &  0.0037  &  0.0155  &  0.0015  &  0.0830  &  0.0010  &  0.0048  &  0.0005  &  0.0302  &  0.0009  &  0.2789  &  0.0056  &  0.0935  &  0.0004  &  \tmult{\nodata}    &  \tmult{\nodata}      \\
%40190  &  \tmult{\nodata}    &  \tmult{\nodata}    &  \tmult{\nodata}    &  0.0018  &  0.0010  &  -0.0021 &  0.0005  &  0.0009  &  0.0005  &  0.0011  &  0.0005  &  0.0078  &  0.0054  &  0.0039  &  0.0023    \\
50001  &  0.0034  &  0.0007  &  -0.0021 &  0.0003  &  0.0052  &  0.0002  &  0.0010  &  0.0002  &  0.0011  &  0.0001  &  0.0188  &  0.0001  &  0.0033  &  0.0000  &  0.0022  &  0.0001  &  0.0017  &  0.0002    \\
50003  &  0.0011  &  0.0006  &  0.0042  &  0.0007  &  0.0158  &  0.0003  &  -0.0004 &  0.0003  &  0.0008  &  0.0001  &  0.0141  &  0.0001  &  0.0028  &  0.0000  &  \tmult{\nodata}    &  \tmult{\nodata}      \\
50006  &  0.0133  &  0.0009  &  0.0064  &  0.0003  &  0.0243  &  0.0003  &  0.0027  &  0.0001  &  0.0049  &  0.0002  &  0.0498  &  0.0007  &  0.0113  &  0.0001  &  \tmult{\nodata}    &  \tmult{\nodata}      \\
50007  &  -0.0010 &  0.0015  &  0.0038  &  0.0006  &  0.0031  &  0.0006  &  0.0004  &  0.0003  &  0.0014  &  0.0002  &  0.0141  &  0.0003  &  0.0039  &  0.0001  &  0.0038  &  0.0001  &  0.0019  &  0.0001    \\
50009  &  0.0025  &  0.0004  &  0.0033  &  0.0003  &  0.0124  &  0.0004  &  0.0001  &  0.0002  &  0.0020  &  0.0001  &  0.0226  &  0.0002  &  0.0064  &  0.0002  &  \tmult{\nodata}    &  \tmult{\nodata}      \\
60003  &  0.0034  &  0.0010  &  -0.0009 &  0.0004  &  0.0007  &  0.0005  &  0.0004  &  0.0003  &  -0.0000 &  0.0001  &  -0.0007 &  0.0001  &  -0.0002 &  0.0002  &  -0.0006 &  0.0002  &  -0.0002 &  0.0001    \\
60006  &  0.0122  &  0.0007  &  0.0029  &  0.0009  &  0.0102  &  0.0005  &  0.0011  &  0.0001  &  0.0089  &  0.0003  &  0.0647  &  0.0010  &  0.0250  &  0.0002  &  \tmult{\nodata}    &  \tmult{\nodata}      \\
60011  &  0.0333  &  0.0004  &  0.0090  &  0.0003  &  0.0210  &  0.0010  &  0.0019  &  0.0002  &  0.0116  &  0.0006  &  0.0926  &  0.0007  &  0.0328  &  0.0000  &  0.0180  &  0.0001  &  0.0114  &  0.0003    \\
\enddata

\tablecomments{Integrated line fluxes are in units of erg\,s$^{-1}$\,cm$^{-2}$ and have been multiplied by a factor of $10^{15}$}.
\end{deluxetable}

Emission line strengths were derived through spectral line indices with appropriate continuum windows (Lick indices). We employ \texttt{IndexF} \citep{cardiel07} to systematically fit \hb, [O\three], [O\one]6300, \ha, [N\two], and [S\two] lines across the entire sample (not only emission-line spectra). \hb\ and [S\two] are not always present due to the distance of some sources from the major axis of observation (as described in Section~\ref{sec:data}). Line fits are listed in Table~\ref{tab:indexf} and emission line ratios will be visited in Section~\ref{sec:excitation}. 

\subsection{Multi-Component Fitting of Broad-Line Spectra}\label{sec:pan}
\begin{figure*}[htbp]
	\begin{center}

		\includegraphics[width=0.6\textwidth,angle=0]{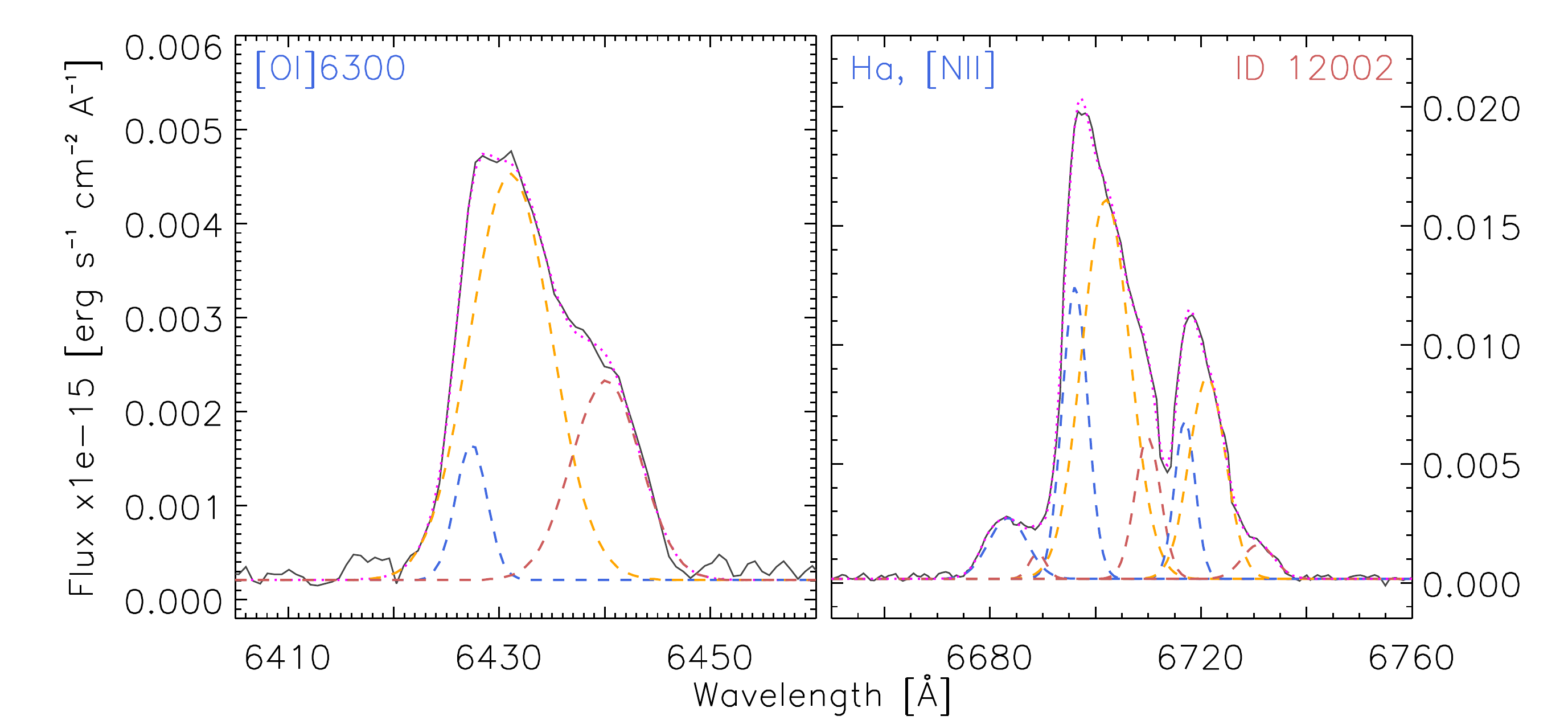}
		\includegraphics[width=0.6\textwidth,angle=0]{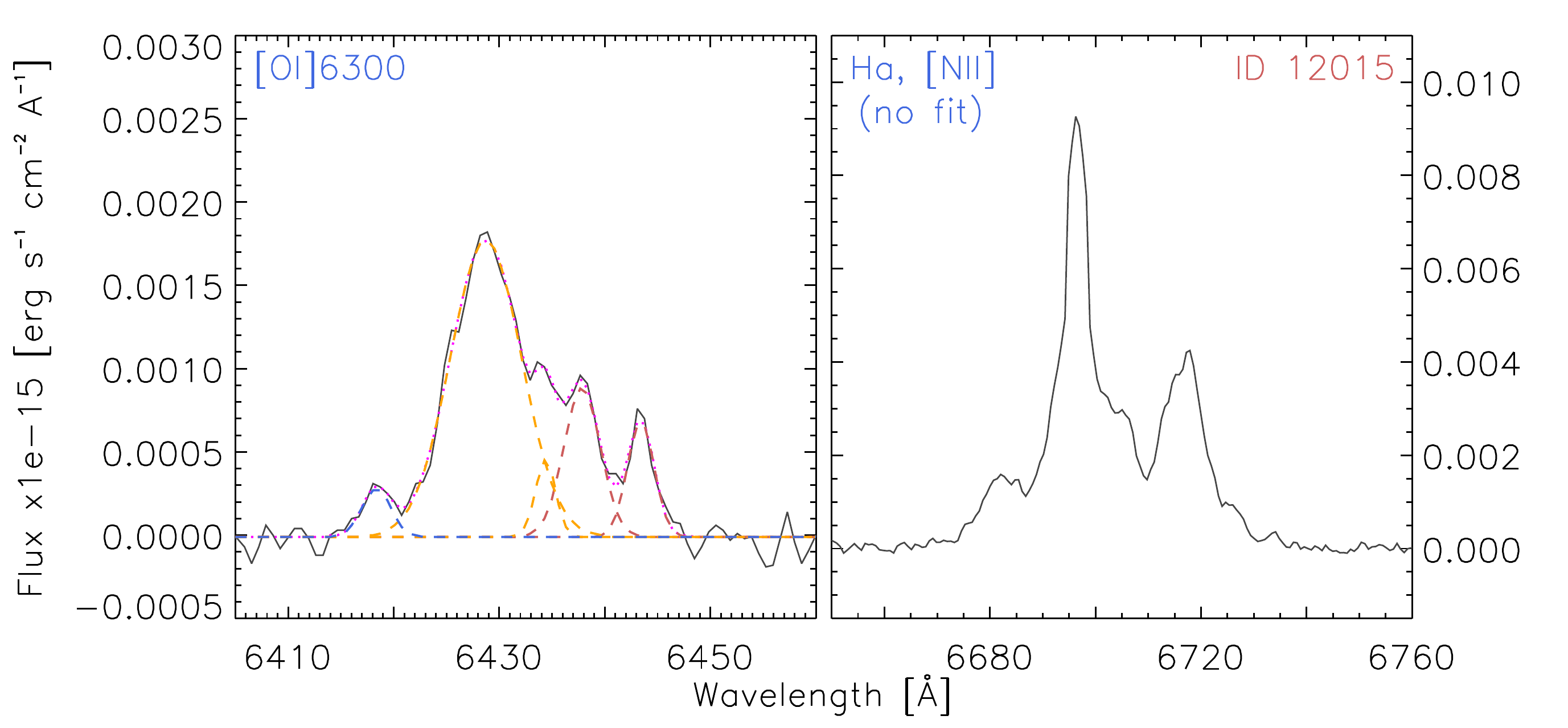}
		\includegraphics[width=0.6\textwidth,angle=0]{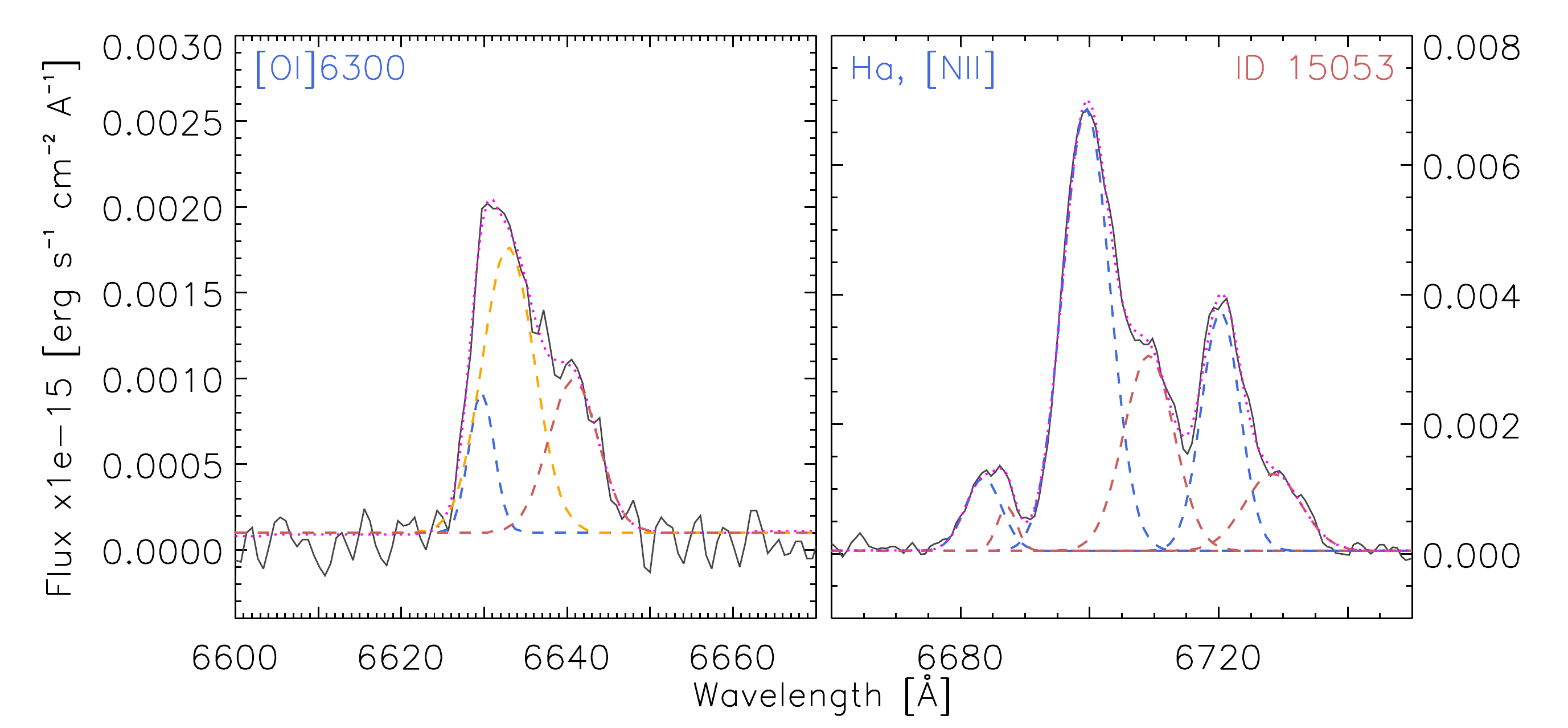}
		\includegraphics[width=0.6\textwidth,angle=0]{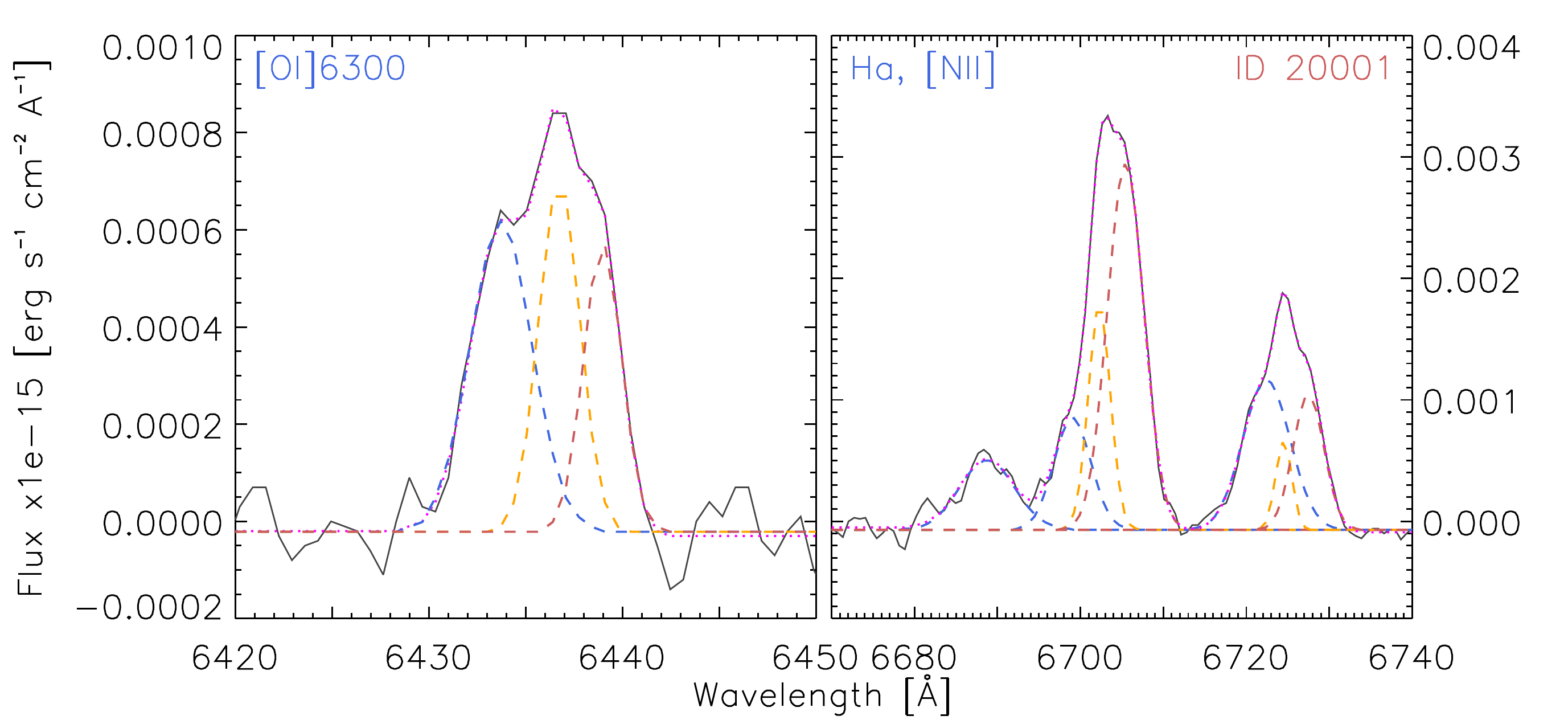}
		\caption{Decomposition of the four broad-line 
			spectra with sufficiently high S/N. A template 
			is constructed based on the [O\one]6300 line 
			before being applied to the \ha/[N\two] region. 
			We typically find a three-component fit to 
			provide a reasonable solution, except in the 
			case of K-12015, where four or more sub-clumps 
			might be present, possibly including a 
			narrow-line feature. The moderate spectral 
			resolution does not allow for further analysis, 
			although we can deduce that all components 
			are broader than the instrumental resolution, 
			and therefore the linewidths of regular 
			photo-ionized \ha\ ($\approx30~$\kms), see text.
		}\label{fig:pan}

	\end{center}
\end{figure*}

%-----------------------------------
Standard methodology, such as that outlined in the previous section, is not appropriate for extracting information from blended spectra. Instead, we used the PAN\footnote{written by Rob Dimeo (NIST) and distributed online at \texttt{http://www.ncnr.nist.gov/staff/dimeo/panweb/pan.html} A version of the software adapted to astronomy can be found at \texttt{http://ifs.wikidot.com/pan} \citep{ifs_wiki}.} algorithm and interface to decompose the broad-line regions, those displaying broadening with respect to the instrumental resolution of $\approx40~$\kms\ \citep[\cf\ typical H\two\ region line widths of 30~\kms;][]{fich90}, into individual emitting knots. This decomposition is shown in Figure~\ref{fig:pan}, and it applies to sources K-12002, K-12015, K-15053, K-20001, and K-20002. We do not plot K-20002, as its spectrum is too faint to trace across the detector. Its broad-line structure is, however, detectable in its two-dimensional spectrum (such as those shown in Figure~\ref{fig:twod}). We also note that the faint K-13011 shows an ambiguous signature that may be interpreted as multi-component or broad emission.

%======= TABLE: PAN =======
\begin{deluxetable}{lcccccc}
\tabletypesize{\normalsize}
\tablewidth{0pt}
\tablecolumns{7}
\tablecaption{%
  Kinematical Decomposition of Broad-Line Spectra.%
  \label{tab:broadline}%
}
\tablehead{%
  \colhead{ID} & 
  \colhead{$u_R(1)$} & 
  \colhead{$u_R(2)$} & 
  \colhead{$u_R(3)$} &
  \colhead{$\sigma(1)$} & 
  \colhead{$\sigma(2)$} & 
  \colhead{$\sigma(3)$} \\
  & 
  \multicolumn{6}{c}{(\ha\ line, \kms)}
}
\startdata
K-12002 &  6122 & 6396 & 6762 & 130 & 244 & 131\\
K-12015 &  \nodata & \nodata & \nodata & \nodata & \nodata & \nodata\\
K-15053 &  6259 & 6716 &  $-$ & 197 & 211 & $-$\\
K-20001 &  6259 & 6396 & 6533 & 120 & 72 & 118
\enddata
\end{deluxetable}
%
%  -ID-         -WAVELENGTH / dispersion-
%IK-12002 &  6696 / 5.67  &  6702 / 10.68  &  6710 / 5.75 \\
%IK-12015 &  \nodata & \nodata & \nodata \\
%IK-15053 &  6699 / 8.61  &  6709 / 9.23  &  \nodata \\
%IK-20001 &  6699 / 5.26  &  6702 / 3.15  &  6705 / 5.17 \\
% 
% sigma = FWHM/lambda * c/instr, where inst~=2.0 A
% 
% Therefore FWHMs are: 
% 5.67, 10.68 -> 130., 244., 131.
% 8.61, 9.23 -> 197., 211.
% 5.26,  3.15, 5.17 -> 120., 72., 118. 
% 

In order to determine the appropriate number of components for the \ha\ line we derive a template from [O\one]6300, as it is not subject to contamination from neighboring features. We fit up to four components to the [O\one] line, as any additional Gaussians are rejected by the algorithm. We place a 2~\AA\ lower limit to the FWHM of each Gaussian matching the instrumental resolution we measured from sky lines. The results of the kinematical decomposition are listed in Table~\ref{tab:broadline}. 

%======= FIGURE: PAN =======

All components fit by the PAN algorithm are significantly broader than the instrumental profile, typically $\approx5~$\AA, suggestive of an excitation process other than photoionization. \hst\ H$\alpha$ imaging (see Figure~\ref{fig:twod}) relates these spectral components to discernible sub-structure. The clumps are, in fact, filamentary, but spatially blended in the ground-based imaging and spectroscopy (the typical seeing is 0\farcs8). Given the positions of these knots in the shock region and ``AGN bridge'', we expect shocks to be the most likely ionizing mechanism, and we will examine this notion in detail in the following section.

%-----------------------------------------------------
\section{Shocked Gas and Star Formation in Stephan's Quintet}\label{sec:star-formation}
%-----------------------------------------------------

\subsection{Velocity Structure of \ha\ Emitting Regions}
\begin{figure*}[tbhp]
	\begin{center}

		\includegraphics[width=0.9\textwidth,angle=0]{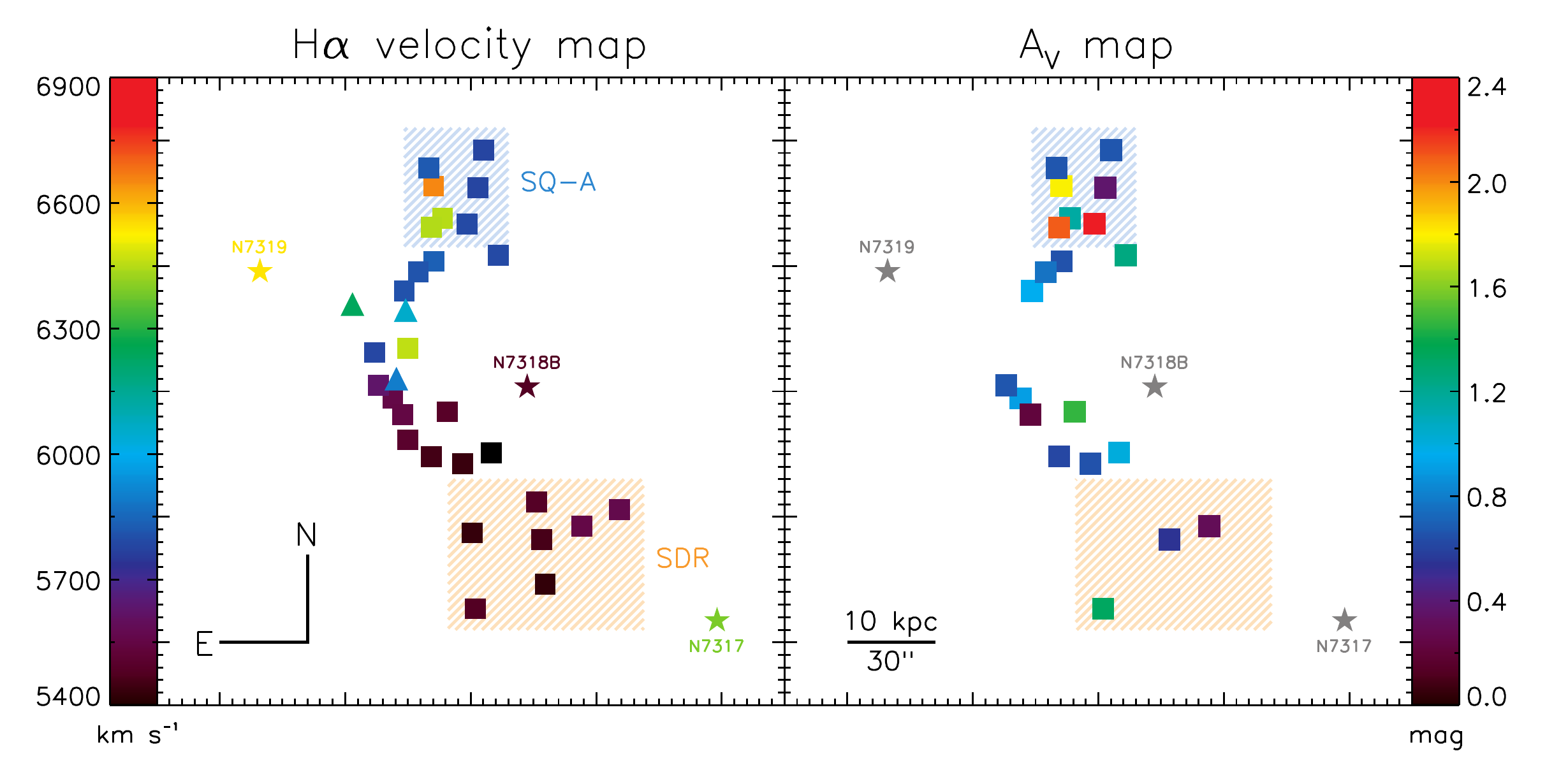}
		\caption{Spatial maps of Stephan's Quintet with measured properties 
			over-plotted and colour-coded %
			\textbf{Left}: \ha\ velocity map of narrow-line
			\ha-emitters in the Quintet as boxes, broad-line knots 
			as triangles (only the dominant component plotted), 
			and galaxies as stars. The scatter of points in 
			the south represents the SDR, at the velocity of 
			the intruder, \n7318B. The arc above is projected 
			across the shock region and displays a northward 
			velocity gradient (see also Figure~\ref{fig:spectra}). 
			North of this point a velocity of 5800~\kms\ is 
			the norm. Besides these coherent regions, three 
			knots (green boxes, the southernmost may be 
			a broad-line knot) are found at the IGM velocity 
			of $\approx6600~$\kms, while one H\two\ region, 
			K-30113 (marked in orange), is found 
			at the high velocity of a recently discovered 
			CO cloud of $\approx6860$~\kms~\citep{guillard12}. 
			\textbf{Right}: Reddening map, including only 
			spectra with high S/N in the H$\beta$ spectral 
			region (galaxies plotted for spatial reference). 
			There are no discernible patterns apart from the 
			high-velocity clumps seen through the relatively 
			high extinction column ($A_V > 1.5$~mag) of the 
			SQ-A starburst region. 
		}\label{fig:velocity}

	\end{center}
\end{figure*}

%-----------------------------------
Figure~\ref{fig:velocity} maps the velocity and reddening measurements of all \ha-emitting knots across an RA-DEC plane representing Stephan's Quintet (north is to the top and east is to the left). We have marked the three galaxies for which we obtained spectra, the SDR, and SQ-A for reference, and the main shock region lies between the two shaded rectangles. We plot only the dominant velocity component of broad-line spectra (as triangles), but omit their extinction in plot. Also, the footprint of each aperture has been exaggerated for illustrative purposes. The velocity map shows significant substructure, with spectra arranged in three phase-space groupings: 

\begin{enumerate}

	\item[(a)] \vspace{-1pt}$\simeq5600~$\kms, the systemic velocity of 
		the intruder, \n7318B. This grouping covers the 
		lower half of the bow structure and the SDR, and 
		hence the majority of \ha-emitting regions projected
		across the large-scale shock. We note a slight, 
		northward velocity gradient throughout. 
	\item[(b)] $\simeq5800~$\kms, the velocity field of the 
		arc to the east of \n7318B. This covers the northern 
		part of the bow structure projected across the shock 
		region and extends north to SQ-A. 
	\item[(c)] 6400--6800~\kms, where we find four clumps: 
		three have velocities consistent with the IGM and 
		the fourth coincides with a CO feature recently 
		discovered by \citet{guillard12}. 

\end{enumerate}

%======= FIGURE: Clump velocity and extinction =======

The corresponding extinction measurements show a largely random distribution, with, perhaps a higher overall extinction in SQ-A. This region, however, shows the highest SFR anywhere in the system \citep{xu03}, which can account for the higher extinction. Note that several data-points do not feature on this plot, as the corresponding sources have either very low S/N ratios or no coverage in the \hb\ region.

\subsection{Excitation Properties and Metallicity}\label{sec:excitation}
\begin{figure}
	\begin{center}

		\includegraphics[width=0.45\textwidth,angle=0]{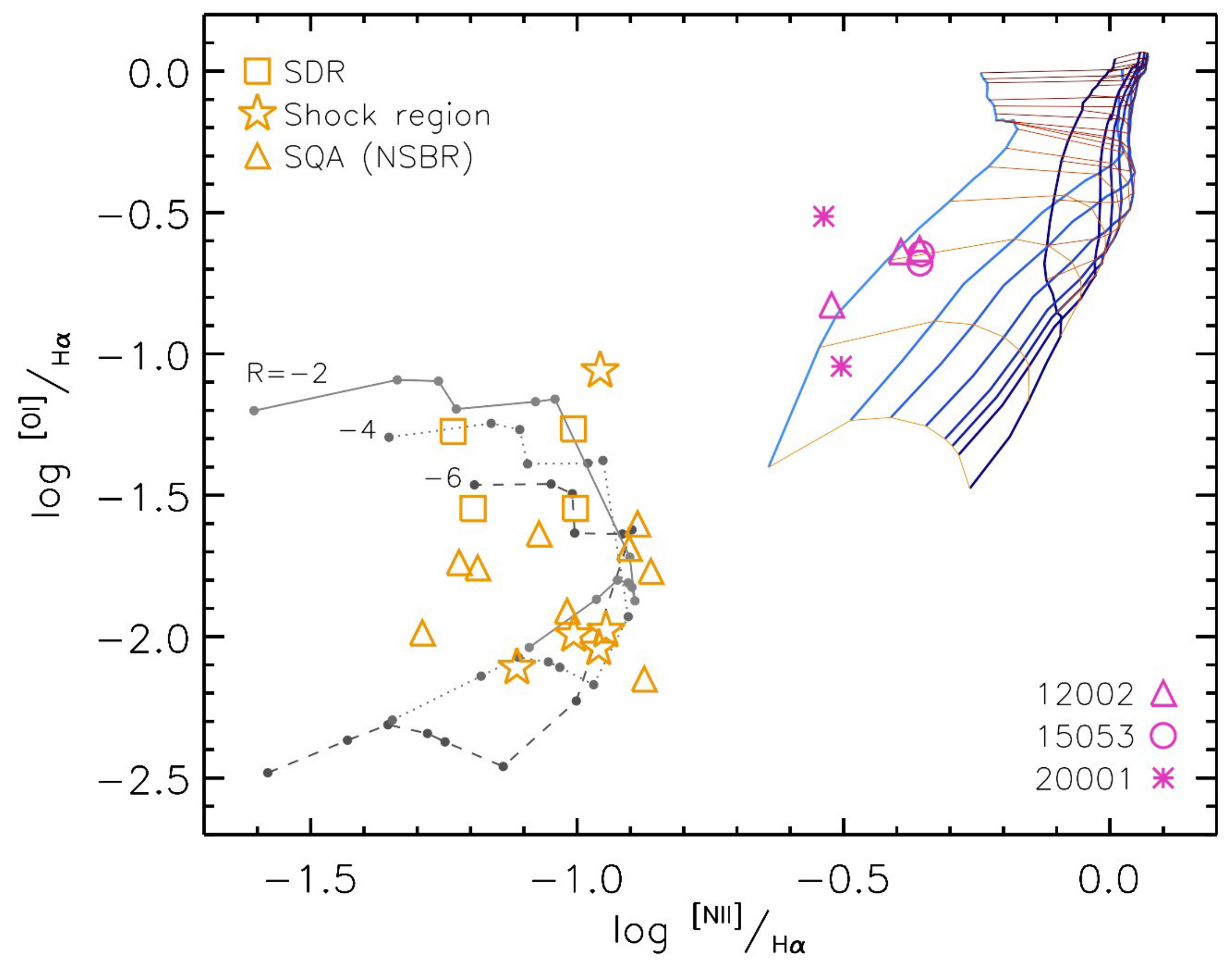}
		\caption{Ionization diagnostics. 
			Orange symbols on the left-hand side mark 
			H\two\ region-like spectra, while magenta 
			symbols on the right mark the decomposed 
			broad-line spectra described in Section
			\ref{sec:pan}. 
			Gray lines mark models of H\two\ region 
			evolution from \citet{dopita06}, with SMC
			metallicity ($0.4\times\Zsun$). 
			The ionization parameter decreases 
			from top to bottom, while time increases 
			clockwise between 0.1 and 6~Myr. 
			The grid on the top right shows a 
			\textit{MAPPINGS} \citep{allen08} 
			shock-plus-progenitor model grid of SMC 
			metallicity ($0.4\times\Zsun$) and an electron density of 
			1~cm$^{-3}$, the combination of parameters 
			found to best fit the data-points. Magnetic 
			field strength increases from left to right, 
			$-3<\log(B/\mu\textup{G})<2$ in irregular increments, 
			while the shock velocity ranges between 200 
			and 1000~\kms\ in 50~\kms\ steps. 
			The distinction between H\two\ regions and shocked
			gas is rather striking and points to a different 
			nature for broad- and narrow-line spectra, rather 
			than a superposition of elements across a continuum
			of velocities. 
		}\label{fig:ionization}

	\end{center}
\end{figure}

%-----------------------------------
The excitation properties of all regions are investigated through comparing two emission line ratios, [O\one]/\ha\ and [N\two]/\ha. We list all line ratios in Table~\ref{tab:ratios}, while Figure~\ref{fig:ionization} plots them against two sets of model tracks: the grayscale lines on the left show models of H\two\ region evolution \citep[photoionization tracks;][]{dopita06} with three different ionization parameters ($R$); while the grid on the right shows a full suite of \textit{MAPPINGS} models \citep{allen08} of shock-plus-progenitor ionization with electron density set to 1~cm$^{-3}$. The magnetic field strength ranges between [0.001, 100]~$\mu$G from left to right and data align with very weak magnetic fields. The shock velocity scales from bottom to top between [200, 1000]~\kms\ in 50~\kms\ intervals, but the plotted line ratios are suggestive of shock velocities no greater than $300~$\kms. This is less than half the velocity of the collision, consistent with the estimate by \citet{cluver10} for the H$_2$-emitting shocked IGM (based on the [Ne\two]/[Ne\three] line ratio). Data-points are region-coded to distinguish between the shock region, SQ-A, and SDR on the left, and the three decomposed broad-line clumps of Section~\ref{sec:pan} on the right (K-12002, K-15053, and K-20001). 

%======= TABLE: Ionisation =======
\begin{table}
\caption{Emission Line Ratios of spectra with valid [O\one] line measurements.}
\begin{center}
	\begin{tabular}{lrr}
	\hline
	\vspace{1pt}ID & 
	$\log\frac{[\textup{\scriptsize N\two}]}{\textup{\scriptsize H}\alpha}$ & 
	$\log\frac{[\textup{\scriptsize O\one}]}{\textup{\scriptsize H}\alpha}$\\
	\hline
	\multicolumn{3}{l}{\textbf{H\two\ regions}}\\
	10006 &    $-$1.01 &    $-$1.99 \\
	11002 &    $-$0.95 &    $-$1.98 \\
	12003 &    $-$0.88 &    $-$1.86 \\
	12004 &    $-$0.87 &    $-$2.15 \\
	12005 &    $-$1.19 &    $-$1.76 \\
	12010 &    $-$1.11 &    $-$2.11 \\
	12013 &    $-$0.96 &    $-$2.04 \\
	12014 &    $-$1.20 &    $-$1.55 \\
	14006 &    $-$0.89 &    $-$1.60 \\
	14030 &    $-$0.96 &    $-$1.06 \\
	30109 &    $-$1.29 &    $-$1.99 \\
	30112 &    $-$1.09 &    $-$2.08 \\
	30113 &    $-$1.22 &    $-$1.74 \\
	30139 &    $-$1.07 &    $-$1.64 \\
	30170 &       0.00 &    $-$4.47 \\
	30174 &    $-$1.02 &    $-$1.91 \\
	50001 &    $-$1.23 &    $-$1.27 \\
	50006 &    $-$1.01 &    $-$1.27 \\
	50007 &    $-$1.00 &    $-$1.55 \\
	60006 &    $-$0.86 &    $-$1.77 \\
	60011 &    $-$0.90 &    $-$1.69 \\
	\multicolumn{3}{l}{\textbf{Decomposed clumps}}\\
	12002~(1) & $-$0.36 &    $-$0.16 \\
	12002~(2) & $-$0.39 &       0.03 \\
	12002~(3) & $-$0.52 &    $-$0.31 \\
	%15053~(1) &  0.00 &    $-$3.51 \\
	15053~(1) & $-$0.36 &       1.42 \\
	15053~(2) & $-$0.35 &       0.98 \\
	20001~(1) &    0.23 &       0.80 \\
	20001~(2) & $-$0.54 &       0.50 \\
	20001~(3) & $-$0.50 &       0.57 \\
	\hline
	\end{tabular}
\end{center}
	%\tablecomments{This Table only lists .}
\label{tab:ratios}
\end{table}
%=======================================

%======= FIGURE: Ionisation plot =======

The metallicity of both model grids in the ionization diagram is set to that of the SMC ($0.4\times\Zsun$), as it is found to provide the best overall fit to the data. This seems to counter the frequent finding of metal-enriched star formation in interacting systems \citep[including certain regions of the Quintet;][]{trancho12}, however, it is consistent with the recent results of \citet{iglesias12}, who measure sub-solar values for the debris-borne H\two\ regions in the Quintet. We do find some clumps consistent with solar metallicity, while the high-velocity clump K-30109 appears to have a lower abundance, perhaps at the 0.2~\Zsun\ level. Lacking strong metallicity tracers, such as the [O\two]3727 line, we cannot confirm the metallicity as tentatively deduced from Figure~\ref{fig:ionization}. A secondary, but unfortunately no more accurate, measure of metallicity is the \citet{kk04} parameterization of the [N\two]/\ha\ ratio. Assuming a low ionization parameter, $q$, following the low electron density, this ratio presents a consistent image of low metallicity throughout the system, with the same few outliers at solar and fifth-solar abundance levels.

\subsection{Star Formation Across the Quintet}\label{sec:SFR}
%---------------------------------------------
Some of the most interesting past findings on the Quintet are those treating the baryon budget within this system. For example, there is more mass locked in a cool, molecular phase than neutral H\one~\citep{cluver10}. In order to take a step closer to completing our understanding of this budget we investigate the rate and efficiency of star formation among our sample of H\two\ regions. 
The extinction-corrected flux of the \ha\ line can be used to derive the SFR of each knot. We disregard sources with low H$\beta$ flux and hence uncertain $A_V$ derivations (see Section~\ref{sec:hiifit}), convert extinction-corrected \ha\ fluxes to luminosities through the inverse-square law \citep[using $d=94$~Mpc for consistency with previous work, \eg][]{guillard12}, and use the \citet{kennicutt98araa} prescription to convert luminosity to an SFR through the relation: 
$$\textup{SFR~[\Msun~yr}^{-1}] = 7.9\times10^{-42}~L_{\textup{\scriptsize H}\alpha}~[\textup{erg~s}^{-1}]$$
We note that \ha luminosities alone will not account for star formation activity within an H\two\ region in its entirety, owing to the geometry-dependent obscuration. Lacking high-resolution infrared imaging we are unable to apply a correction for obscuration, meaning that our SFR derivations slightly underrepresent the true rates. The total resulting SFR in the 24 qualifying H\two\ regions amounts to 0.084~\Msun\,yr$^{-1}$, more than half of which is found in the two brightest star-forming regions, K-10006 (at the southern edge of the shock region) and K-30109 (the central source in SQ-A). 
As described in Section~\ref{sec:completeness}, our sample is biased toward the most luminous (massive) H\two\ regions in Stephan's Quintet. It is therefore not advisable to extrapolate SFR values over the entirety of the shock region or SQ-A. 
We can, however, contrast the SFR of our sources with the output of star-forming regions in general. In order to place these values in the context of regular star-forming regions we perform two tests. 

First we compare to nearby star-forming regions. For this we need to note that our spectra represent light extracted over 0.75~sq.~arcsec to reflect typical seeing, although most knots are intrinsically smaller. At the 94~Mpc distance to the Quintet, that is a physical diameter of $\approx450~$pc, comparable to large star-forming regions in nearby galaxies---\eg\ the giant H\two\ regions \n604 \citep[$d\approx270~$pc;][]{jesus04} and especially 30~Dor \citep[$d\approx300~$pc;][]{evans11}, which is often used as the local benchmark for dense regions of intense star formation activity. Comparing the \ha\ luminosities of our knots directly with 30~Dor we find none to register the same order of magnitude. For the two brightest knots, K-10006 and K-30109, we record luminosities of 0.65 and 0.33 in 30~Dor units, however, the majority of knots register between 0.01--0.10. This comparison indicates that, apart from the two brightest knots, we are looking at unremarkable star-forming regions and complexes. 

For the second test we use the H$_2$ mass measurements of \citet{guillard12} to place our median H\two\ regions on the \citet{bigiel08} formulation of the Schmidt-Kennicutt relation (the `star formation law'). We employ the approximation of a uniform H$_2$ distribution across the \citeauthor{guillard12} IRAM 30m~telescope beams and establish a median H$_2$ mass per velocity component of $5\times10^8$~\Msun, over a beam area of 547 sq.\ arcsec. We therefore estimate an H$_2$ surface density of $6.9\times10^5$~\Msun\ per slit (0.75~sq.\ arcsec or 0.12~kpc$^{2}$), which translates to $\log(\Sigma_{\textup{\scriptsize H}2})=0.77$~\Msun~pc$^{-2}$. The typical SFR among our sample, which, owing to the spectroscopic selection represents the brighter end of the luminosity function, is $10^{-3}~$\Msun\ yr$^{-1}$, or $\log(\Sigma_\textup{\scriptsize SFR})=-2.07~$\Msun~yr$^{-1}$~kpc$^{-2}$, and these two values place the average \ha-emitter slightly above the main locus of the \citet{bigiel08} data-points, at a star formation efficiency of 10\%. Considering that 20\% is considered an upper limit \citep[\eg,][]{norman11}, this is relatively high compared to regular star-forming regions in nearby galaxies. Note, however, that the uncertainty of the uniform approximation may shift the datapoint to higher H$_2$ content, and therefore on or below the locus (as the SFR measurement remains fixed). Lacking spatial information on the CO beams, it is not possible to assess the efficiency of the brightest \ha-emitters in our sample, as the approximation should not be expected to hold.

%-----------------------------------------------------
\section{Shock-Suppression, CO cooling, and Star Formation}\label{sec:discussion}
%-----------------------------------------------------
The results presented so far have split the \ha-emitting regions into two classes: regular H\two\ regions, and clumpy filaments of shocked molecular gas. Having derived the kinematical structure, ionization characteristics, metallicity and extinction distributions of all \ha\ emitters, along with star formation rates and masses for the H\two\ regions, we can now assess the role of shocks in affecting the star formation process in Stephan's Quintet. 

% ======= FIGURE: Hii region CC plot =======
\begin{figure*}[tbhp]
	\begin{center}

		\includegraphics[width=0.8\linewidth,angle=0]{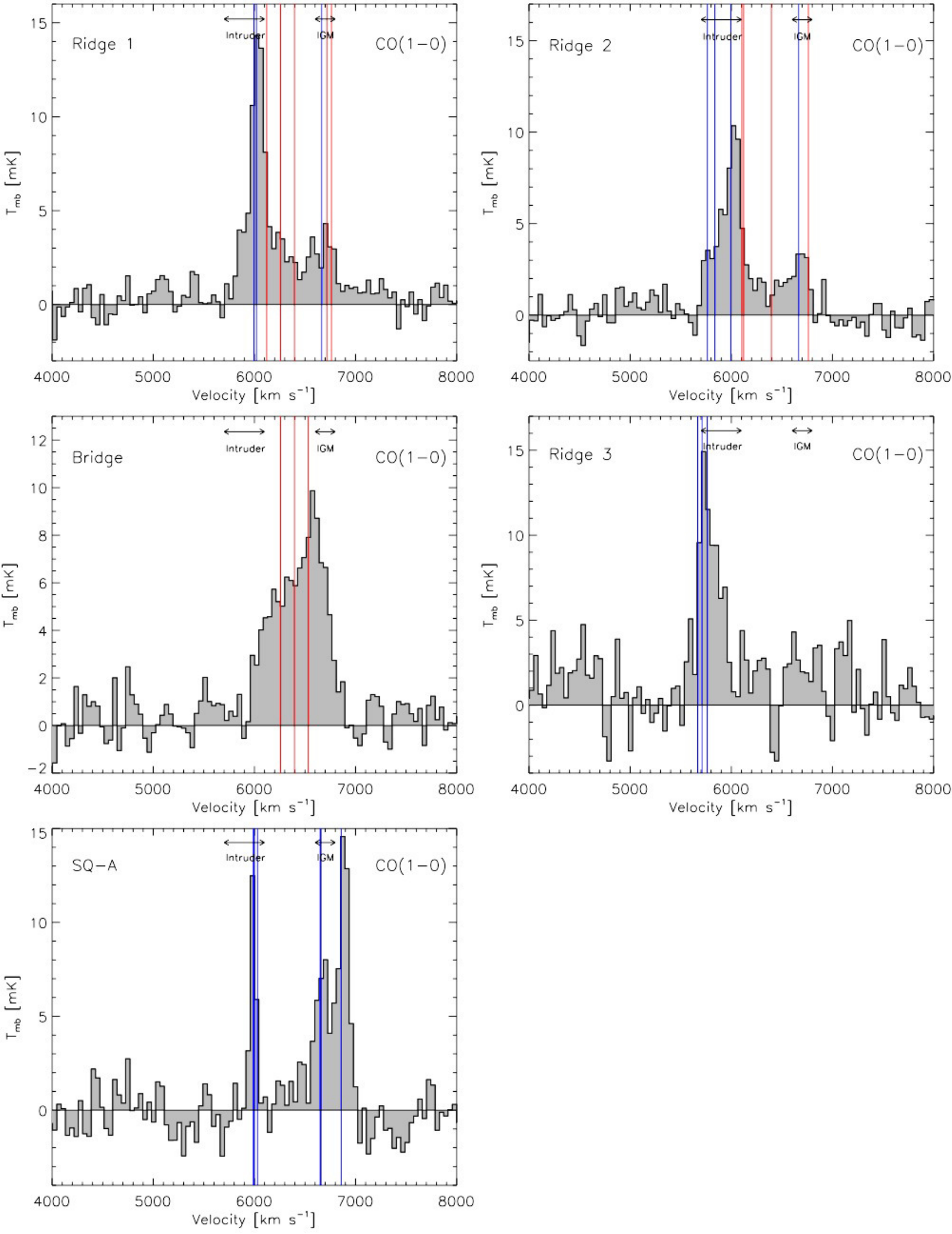}
		\caption{%
			Velocity diagrams of molecular gas from 
			\citet[][as indicated in Figure~\ref{fig:finder}]{guillard12}, 
			with those \ha-emitters within 40\% of each 
			CO beam-width indicated as blue and red lines for 
			H\two\ regions and shocked clumps respectively. 
			Ridges 1, 2, and 3 cover the shock region 
			in three consecutive, north-to-south pointings. 
			The slight overlap between Ridges~1 and 2 leads 
			to the duplication of some red lines. The shocked 
			clumps in the ``AGN bridge'', corresponding to the 
			decomposed knots in K-20001, line up with the 
			observed CO emission but not the IGM, nor the 
			intruder. The majority of H\two\ regions are aligned 
			with the intruder and also match the high-velocity 
			CO clouds projected against SQ-A. In Ridges~1 
			and 2, which coincide with the peak of X-ray 
			emission, we see a mix of H\two\ regions and 
			shocked filaments in the IGM velocity, but not 
			in the intruder, where there is only star formation. 
			The sharp boundary between the CO distribution 
			of the intruder and the velocity of shocked clumps 
			indicates that material in \n7318B lies 
			behind the shock in our line of sight, and 
			hence able to condense into star forming regions. 
		}\label{fig:CO}

	\end{center}
\end{figure*}

Given the complex nature of the IGM the two classes of \ha-emitters need to be placed in the context of the multiphase gas. To facilitate this comparison, we present in Figure~\ref{fig:CO} the velocity distribution of cool molecular gas, diagnosed through the CO(1-0) line in five pointings, from \citet{guillard12}. To this we have superposed the velocities of all spatially coincident \ha-emitters. 

From this it is evident that the shocked filaments span a large range in redshift, inconsistent with the frame of reference of the intruder. The shock velocities derived for these filaments agree with those for the shock-wide H$_2$ emission, which we know to be the dominant emission component \citep{appleton06,cluver10}. It is therefore reasonable to associate the broad-line emitters with the warm H$_2$ emission. The typically smaller velocity they display than the IGM indicates deceleration from the oncoming shock front, consistent with the turbulent dissipation of kinetic energy modeled by \citet{guillard12}. 

The kinematics of the H\two\ regions suggest a different distribution in three-dimensional space than the warm molecular filaments. While in SQ-A they appear associated with both the intruder and IGM velocity brackets, across the shock region they are found exclusively in the frame of reference of the intruder. The only possible exception, an H\two\ region in the shocked IGM, is K-13011, a faint, highly extinguished and compact source of ambiguous nature -- its signal-to-noise ratio does not allow for line decomposition and it is likely to be shock-broadened. Based solely on high-confidence detections, our conclusion is that high-mass star formation is entirely inhibited by the shocks throughout the IGM of Stephan's Quintet. Since our observations do not cover low-mass H\two\ regions, however, it is possible that star formation may proceed at lower levels, undetected by our observing campaign. 

Overall, the kinematics of the H\two\ regions projected against the shock and their relatively low masses indicate that they are part of the leading spiral arm of the intruder, or perhaps a tidally-induced outer edge pileup, akin to those arising in the dynamical models of \citet{struck12}. The slight, northward velocity gradient strengthens this scenario, as it can be attributed to the (assumed prograde) rotating motion of the galaxy. In addition, the considerable reddening (median $>1~$mag) of the star-forming regions indicates that we may be peering at them through a volume of intra-group material. This interpretation is consistent with the slight spatial mismatch between PAH features and X-ray emission, as discovered by \citet{cluver10}. The sharp boundary seen in Figure~\ref{fig:CO} between CO in the intruder and the red lines indicating shocked clumps suggests that the material within the intruder has not been subjected to heating in the same way as the IGM has. The lopsided distribution of \ha\ emitters in the intruder might indicate that a starburst has yet to occur, as might be expected from the timescale of this interaction---\cf\ the interacting pairs of \citet{scudder12}, where the SFR peak often occurs as an interactor travels toward apogalacticon.

The marked difference between the main shock region and SQ-A, where star formation does not appear to be inhibited, is likely due to the longer period over which SQ-A has been forming stars. The previous generation of interaction events sparked the current starburst in this locale, as testified by the ages of star clusters \citep{fedotov11}, which are not restricted the embedded phase ($\tau_\textup{\scriptsize age}\lesssim10~$Myr). \citet{cluver10} and \citet{guillard12} find comparable masses of warm and cool molecular gas in SQ-A, implying not that the gas was never heated, but that the SF process was well underway when the shock propagated through the medium some time in the past 10~Myr \citep[or perhaps that the dissipation timescale is shorter in SQ-A;][]{guillard09,guillard12}. 

Finally, it is worthwhile considering Stephan's Quintet in the context of its projected evolution into a set of early-type galaxies with a bright X-ray halo. This will place it in Sequence~B of the \citet{isk10} evolutionary diagram, where CGs evolve within a shared IGM. Known examples of this class include HCG~42, whose star formation history displays a striking, abrupt cliff near the 2~Gyr mark \citep{isk13b}. Such a star formation history is perhaps to be expected in the Quintet, although, given the many participants in the Quintet interaction and the multiple ongoing starbursts, the quenching of star formation brought on by this shock might not be as clear-cut as the situation in HCG~42 \citep[see the clumpy color-color plots of][]{fedotov11}. 

%-----------------------------------------------------
\section{Summary}\label{sec:summary}
%-----------------------------------------------------
We have presented an investigation of the interplay between the strong, extended shock in Stephan's Quintet and the formation of new stellar associations across this multi-galaxy system. In order to understand the role of the shock in regulating the star formation process, we used Gemini GMOS-N to obtain high-quality optical spectra of $\approx40$ \ha-bright clumps across this compact galaxy group, covering the spectral range between H$\beta$ and [S\two]. The sources were selected on \hst\ images, giving us the ability to map the excitation in this shock region at an unprecedented level of detail. 

The studied \ha\ emitters display a bimodality in terms of their emission-line widths and ionization parameters. We find a set of five (possibly six) broad-line objects, which we decompose into several (three or four) individual clumps. These sources are filamentary in nature, as evident in \ha\ imaging from \hst, and each individual knot exhibits broadening at a multiple of the 40~\kms\ resolution element---with the possible exception of a single narrow component in K-12015, which we were unable to decompose. By studying their ionization characteristics, we identify all components that make up these filaments as molecular gas heated by the propagation of a shock front at $100-300~$\kms, consistent with the H$_2~0$-0~S(1) emission mapped by \citet[][which traces the X-ray emitting medium]{cluver10}. While these H$_2$ clumps are significantly cooler than the X-ray emitting gas that surrounds them, they display no evidence of star formation. 

The population of $\approx\,$30~H\two\ regions show SFRs and SFR-densities similar to nearby, regular star-forming regions, with the exception of two knots that are more akin to prolific stellar nurseries such as 30~Dor. As a whole, they can be split into three sub-components according to their spatial distribution: those in the southern debris region (SDR), those projected against the shock region, and those being formed in the SQ-A starburst. Kinematically, however, they are more consistent than that. \ha\ emitters in the SDR and shock region display a mean velocity of \mbox{$\approx\,$5600~\kms} and a slight but uninterrupted velocity gradient. This not only places them in the frame of reference of the intruder, but is consistent with the prograde circular motion of a spiral arm. Further support for this thesis stems from several other pieces of evidence: relatively high reddening, a low median SFR, and a striking accord between the spatial distributions of H\two\ regions and PAH features studied by \citet[][with $\approx1\arcsec$ spatial resolution, similar to our GMOS data]{cluver10}, so we associate the star-forming regions with the intruder. \ha-emitters projected against the shock region appear to be situated in a spiral arm of \n7318B. Those in the SDR might be associated with an extension of the spiral arm, a debris structure, although the projected spatial separation between \n7318B and the boundary of the SDR is substantial. Knots in SQ-A are simple products of an ongoing starburst whose onset pre-dates the galaxy-IGM collision that gave rise to the shock front. These pieces of evidence, along with the slightly higher SQ-A velocity are consistent with the \citet{renaud10} and \citet{hwang12} models that trace the origin of SQ-A as debris ejected from \n7318A before colliding with the IGM. Having associated the entire population of H\two\ regions projected against the shock to the intruder instead, we find no evidence for star formation in the shocked IGM. We therefore conclude that the process is altogether suppressed throughout the hot phase of the clumpy medium. 

We find three star-forming regions and one potential broad-line clump outside the above-mentioned phase-space distribution. They are projected across the shock region and SQ-A but lie at higher velocities ($6400-6600~$\kms), in accord with systemic velocity of the IGM. One exception, at $\approx6800~$\kms, is an optical counterpart to a CO cloud recently discovered by \citet{guillard12} and displays a metallicity lower than the average---which is already low at SMC levels. We also present the optical spectrum of \n7318A for the first time, and classify it as an early-type galaxy, confirming the morphological classification of \citet{sulentic95}. 

Understanding the effect of strong shocks on  the evolution of compact galaxy groups is of great importance to the empirical framework of galaxy evolution. The majority of galaxies and stellar mass in the local Universe are found in groups \citep{small99,eke05}, where high-speed collisions are infrequent, but not rare---especially in CGs \citep{cluver13}. The build-up of X-ray haloes in low-mass CGs is conditional on such collisional heating of the IGM, as their shallow potential wells close the conventional channels that dominate in more populous groupings (\eg, the frictional energy exchange between massive galaxies and the rich IGM in clusters). In addition, groupings of galaxies at higher redshifts, before matter assembled to its current stage, might have resembled present-day CGs. If the Quintet is an appropriate interpretive guide to the effect of shock-heating, we can deduce that large volumes of molecular gas can be locked in this hot phase for $\sim1~$Gyr or more at a time, while the group evolves inside an X-ray halo. In that way, the stochastic occurrence of an intruder will define the evolution of a CG---at least the lower branch of the \citet{isk10} evolutionary pathway where strong interactions between gas-rich galaxies dominate. This must be taken into account in models of galaxy evolution and in compiling the baryon budgets of galaxy groups.

\acknowledgements We thank the anonymous referee for helping us reinforce some of the analysis.

Based on observations obtained at the Gemini Observatory (Program ID GN-2010B-Q-56), which is operated by the Association of Universities for Research in Astronomy, Inc., under a cooperative agreement with the NSF on behalf of the Gemini partnership: the National Science Foundation (United States), the National Research Council (Canada), CONICYT (Chile), the Australian Research Council (Australia), Minist\'erio da Ci\^encia, Tecnologia e Inova\c{c}\~ao (Brazil) and Ministerio de Ciencia, Tecnolog\'ia e Innovaci\'on Productiva (Argentina).

Based on observations made with the NASA/ESA Hubble Space Telescope, obtained from the data archive at the Space Telescope Science Institute. STScI is operated by the Association of Universities for Research in Astronomy, Inc. under NASA contract NAS 5-26555

ISK acknowledges support from grant no.\ 1439092 from the Jet Propulsion Laboratory. ISK is grateful for the hospitality of the staff at the NASA Herschel Science Center, the Giant Magellan Telescope Organisation, and Carnegie Observatories, all in Pasadena, CA, where much of this work was undertaken. 

SCG and KF thank the Canadian Natural Science and Engineering Research Council and the Ontario Early Researcher Award Program for support. \bibliographystyle{apj}
\bibliography{references}

\end{document}